\documentclass[preprint]{aastex}
\usepackage{natbib}

\singlespace

\shortauthors{Kogut et al.}
\shorttitle{WMAP TE Polarization}
\slugcomment{Accepted by the \emph{Astrophysical Journal}}

\begin{document}

\title{Wilkinson Microwave Anisotropy Probe ({\sl WMAP}) 
First Year Observations: TE Polarization}

\author{A. Kogut\altaffilmark{2},
D. N. Spergel\altaffilmark{3},
C. Barnes\altaffilmark{4},
C. L. Bennett\altaffilmark{2},
M. Halpern\altaffilmark{5},
G. Hinshaw\altaffilmark{2}, 
N. Jarosik\altaffilmark{4},  
M. Limon\altaffilmark{2,4,6}, 
S. S. Meyer\altaffilmark{7},
L.Page\altaffilmark{3},
G. S. Tucker\altaffilmark{2,6,8},
E. Wollack\altaffilmark{2},
E. L. Wright\altaffilmark{9}
}

\altaffiltext{1}{{\sl WMAP} is the result of a partnership between 
                 Princeton  University and NASA's Goddard Space Flight Center. 
                 Scientific guidance is provided by the 
                 {\sl WMAP} Science Team.}
\altaffiltext{2}{Code 685, Goddard Space Flight Center, Greenbelt, MD 20771}
\altaffiltext{3}{Dept of Astrophysical Sciences, Princeton University, Princeton, NJ 08544}
\altaffiltext{4}{Dept. of Physics, Jadwin Hall, Princeton, NJ 08544}
\altaffiltext{5}{Dept. of Physics and Astronomy, University of British Columbia, Vancouver, BC  Canada V6T 1Z1}
\altaffiltext{6}{National Research Council (NRC) Fellow}
\altaffiltext{7}{Depts. of Astrophysics and Physics, EFI and CfCP, University of Chicago, Chicago, IL 60637}
\altaffiltext{8}{Dept. of Physics, Brown University, Providence, RI 02912}
\altaffiltext{9}{UCLA Astronomy, PO Box 951562, Los Angeles, CA 90095-1562}

\email{Alan.J.Kogut@nasa.gov}

\begin{abstract}

The Wilkinson Microwave Anisotropy Probe ({\sl WMAP}) 
has mapped the full sky
in Stokes $I$, $Q$, and $U$ parameters
at frequencies 23, 33, 41, 61, and 94 GHz.
We detect correlations
between the temperature and polarization maps
significant at more than 10 standard deviations.
The correlations are inconsistent with instrument noise
and are significantly larger than the upper limits
established for potential systematic errors.
The correlations are present in all {\sl WMAP} frequency bands
with similar amplitude from 23 to 94 GHz,
and are consistent with a superposition
of a CMB signal with a weak foreground.
The fitted CMB component is robust
against different data combinations
and fitting techniques.
On small angular scales
$(\theta < 5\arcdeg)$, 
the {\sl WMAP} data 
show the temperature-polarization correlation
expected from adiabatic perturbations in the temperature power spectrum.
The data for $\ell > 20$ agree well with the signal predicted
solely from the temperature power spectra,
with no additional free parameters.
We detect excess power on large angular scales
($\theta > 10\arcdeg$)
compared to predictions based on the temperature power spectra alone.
The excess power is well described by 
reionization at redshift 
$11 < \ensuremath{z_r} < 30$ 
at 95\% confidence,
depending on the ionization history.
A model-independent fit to reionization optical depth
yields results consistent with the best-fit $\Lambda$CDM model,
with best fit value 
\ensuremath{\tau = 0.17 \pm 0.04}
at 68\% confidence, including systematic and foreground uncertainties.
This value is larger than expected
given the detection of a Gunn-Peterson trough
in the absorption spectra of distant quasars,
and implies that 
the universe has a complex ionization history:
{\sl WMAP} has detected the signal from an early epoch of reionization.
\end{abstract}

\keywords{cosmic microwave background, 
cosmology: observations,
instrumentation: polarimeters }

\clearpage

\newpage

\section{INTRODUCTION}

Linear polarization of the cosmic microwave background (CMB)
results from anisotropic Thomson scattering of CMB photons by free electrons.
By symmetry, an isotropic radiation field 
can not generate a net polarization.
Any net polarization results from the quadrupole moment
of the CMB temperature distribution
seen by each scatterer.
Multiple scattering suppresses polarization
by damping the temperature anisotropy;
hence, CMB polarization originates primarily from epochs 
when the opacity was of order unity or less.
Standard cosmological models
predict two such epochs,
corresponding to two characteristic angular scales.
The first is the decoupling surface at redshift $z \approx 1089$,
when the ionization fraction $x_e$ abruptly falls from near unity to near zero.
The acoustic horizon at decoupling
subtends an angle $\theta \approx  1\arcdeg$;
polarization on these scales 
reflects conditions in the photon-baryon fluid
just prior to recombination.
Polarization data from decoupling complement
measurements of the temperature anisotropy.
Astrophysical sources generate additional polarization
as ionizing radiation from the first collapsed objects 
reionizes the intergalactic medium.
For reionization at redshift $z < 50$
the horizon is on large angular scales,
$\theta > 5 \arcdeg$.
Polarization on these scales
directly probes the poorly-understood
process of reionization.

Since CMB polarization originates at modest opacity,
the underlying temperature anisotropy is not heavily damped
and remains observable today.
Precise predictions can be made
of the average polarization pattern expected
from a given power spectrum of temperature anisotropy
(\citet{
rees:1968,
kaiser:1983,
bond/efstathiou:1984,
coulson/crittenden/turok:1994,
kamionkowski/kosowsky/stebbins:1997,
zaldarriaga/seljak:1997,
hu/white:1997};
for recent reviews, see 
\citet{kosowsky:1999,
hu/dodelson:2002}).
The pattern of polarization on the sky is a vector field
with both an amplitude and direction at each point,
and can be separated into 
two scalar fields,
one giving the curl and the other the gradient component
(called $B$ and $E$ modes in analogy with electromagnetic fields).
The DASI collaboration has detected
CMB polarization on angular scales $\sim 0\fdg5$
\citep{kovac/etal:2002}.
DASI reports an $E$ mode signal significant at 4.9$\sigma$
and a TE temperature-polarization correlation
significant at 2$\sigma$.
Both signals are consistent with the
``concordance'' cosmological model
(spatially flat model dominated by a cosmological constant
and cold dark matter;
see, e.g., \citet{hu/dodelson:2002})
and support an adiabatic origin for the CMB temperature anisotropy.

The Microwave Anisotropy Probe has mapped the full sky
in the Stokes $I$, $Q$, and $U$ parameters
on angular scales $\theta > 0\fdg2$
in 5 frequency bands centered at 23, 33, 41, 61, and 94 GHz
\citep{bennett/etal:2003}.
{\sl WMAP} was not designed solely as a polarimeter,
in the sense that none of its detectors
are sensitive only to polarization.
Incident radiation in each differencing assembly (DA)
is split by an orthomode transducer (OMT) 
into two orthogonal linear polarizations
\citep{page/etal:2003,
jarosik/etal:2003}.
Each OMT is oriented so that 
the electric field directions 
accepted in the output rectangular waveguides
lie at $\pm 45\arcdeg$ 
with respect to the $yz$ symmetry plane of the satellite
(see \citet{bennett/etal:2003} Fig. 2
for the definition of the satellite coordinate system).
The two orthogonal polarizations from the OMT
are measured by two independent radiometers.
Each radiometer differences the signal 
in the accepted polarization
between two positions on the sky
(the A and B beams),
separated by $\sim 140\arcdeg$.

\begin{figure}
\epsscale{0.6}
\plotone{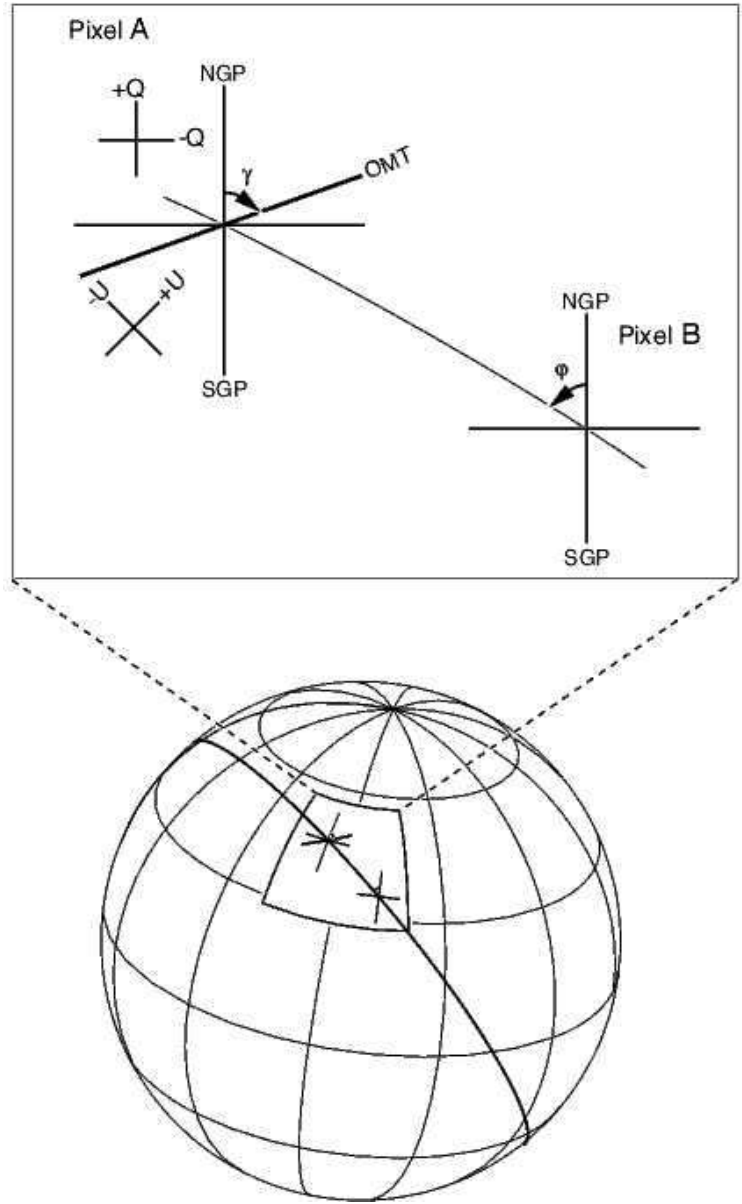}
\caption{Geometry for Stokes $Q$ and $U$ parameters.
{\sl WMAP} measures polarization by 
differencing two orthogonal polarization channels,
then solving for $Q$ and $U$ 
as the spacecraft compound spin
projects the OMT onto the sky 
at different angles $\gamma$
relative to the Galactic meridians.
All analysis uses coordinate-independent quantities
$Q^\prime$ and $U^\prime$
defined with respect to
the great circle connecting a pair of pixels
(see text).}
\label{qu_cartoon}
\end{figure}

The signal from the sky in each direction $\hat{n}$ 
can be decomposed into the Stokes parameters
\begin{equation}
T(\hat{n}) = I(\hat{n}) + Q(\hat{n}) \cos{2 \gamma} + U(\hat{n}) \sin{2 \gamma},
\label{stokes_def}
\end{equation}
where we define the angle $\gamma$ 
from a meridian through the Galactic poles
to the projection on the sky of the E-plane of each output port of the OMT
(Fig. \ref{qu_cartoon}).
In principle,
by tracking the orientation of the OMTs on the sky
as the satellite scan pattern observes each sky pixel
in different orientations,
each radiometer could independently produce a map of the
Stokes $I$, $Q$, and $U$ parameters.
In practice, the non-uniform coverage of $\gamma$ at each pixel
would generate significant correlations 
between the fitted Stokes parameters,
allowing leakage of the dominant temperature anisotropy
into the much fainter polarization maps.
We avoid this problem
by differencing the outputs 
of the two radiometers in each differencing assembly
in the time-ordered data.
Denoting the two radiometers by subscripts 1 and 2,
the instantaneous outputs are
\begin{eqnarray}
\Delta T_1 & = & I(\hat{n}_A) + Q(\hat{n}_A) \cos 2\gamma_A + U(\hat{n}_A) \sin 2\gamma_A \nonumber \\
           & - & I(\hat{n}_B) - Q(\hat{n}_B) \cos 2\gamma_B - U(\hat{n}_B) \sin 2\gamma_B
\end{eqnarray}
and
\begin{eqnarray*}
\Delta T_2 & = & I(\hat{n}_A) - Q(\hat{n}_A) \cos 2\gamma_A - U(\hat{n}_A) \sin 2\gamma_A \\
           & - & I(\hat{n}_B) + Q(\hat{n}_B) \cos 2\gamma_B + U(\hat{n}_B) \sin 2\gamma_B .
\end{eqnarray*}
The sum
\begin{equation}
\Delta T_{I} \equiv \frac{1}{2}(\Delta T_1 + \Delta T_2) = I(\hat{n}_A) - I(\hat{n}_B)
\label{sum_def}
\end{equation}
is thus proportional to the unpolarized intensity,
while the difference
\begin{equation}
\Delta T_{P} \equiv \frac{1}{2}(\Delta T_1 - \Delta T_2)
  = Q(\hat{n}_A) \cos 2\gamma_A + U(\hat{n}_A) \sin 2\gamma_A 
  - Q(\hat{n}_B) \cos 2\gamma_B - U(\hat{n}_B) \sin 2\gamma_B.
\label{diff_def}
\end{equation}
is proportional only to the polarization.
We produce full-sky maps of the Stokes $I$, $Q$, and $U$ parameters
from the sum and difference time-ordered data
using an iterative mapping algorithm.
Since the polarization is faint,
the $Q$ and $U$ maps are dominated by instrument noise
and converge rapidly
\citep{hinshaw/etal:2003b}.

The Stokes $Q$ and $U$ components depend on 
a specific choice of coordinate system.
For each pair of pixels,
we define coordinate-independent quantities
\begin{eqnarray}
 & & Q^\prime = Q \cos(2 \phi) + U \sin(2 \phi) \nonumber \\
 & & U^\prime = U \cos(2 \phi) - Q \sin(2 \phi),
\label{prime_def}
\end{eqnarray}
where the angle $\phi$ rotates the coordinate system 
about the outward-directed normal vector
to put the meridian along the great circle connecting
the two positions on the sky
\citep{kamionkowski/kosowsky/stebbins:1997,
zaldarriaga/seljak:1997}.
All of our analyses use these coordinate-independent
linear combinations of the $Q$ and $U$ sky maps.

Simulations of the mapping algorithm
demonstrate that {\sl WMAP}
can accurately recover the polarization pattern on the sky,
even after allowing for residual calibration uncertainty
in the individual radiometer channels.
However, non-ideal instrumental signals
affect the $Q$ and $U$ sky maps
to a greater extent than the unpolarized $I$ maps.
The spacecraft spin about its $z$ axis
sweeps the beams across the sky
in a direction $45\arcdeg$ from the OMT orientation,
preferentially coupling 
signals not fixed on the sky
into the $U$ map.
Residual striping exists to a lesser extent in the $I$ and $Q$ maps.
Systematic errors in the individual $Q$ and $U$ maps
are not yet fully assessed;
consequently, we defer detailed analysis 
of the $Q$ or $U$ maps 
to a later paper.
Cross-correlations between maps
are largely unaffected by striping
or any other channel-specific signal,
allowing much simpler analysis of the faint polarization signal
than would be possible for the individual $Q$ or $U$ maps.
This paper discusses the 
temperature-polarization (TE) correlation
in the {\sl WMAP} one-year sky maps.

We compute the temperature-polarization cross-correlation
using three different techniques:
the two-point correlation function,
a quadratic estimator for the power spectrum,
and a ``template'' comparison in pixel space
between the polarization maps
and the predicted polarization 
given the observed pattern of temperature anisotropy.
All three methods yield similar results
despite disparate treatments of the data.

\section{CORRELATION FUNCTION}

The simplest measure of temperature-polarization cross-correlation
is the two-point angular correlation function
\begin{equation}
C^{IQ}(\theta) = \frac
{\sum_{ij} I_i Q^\prime_j w_i w_j }
{\sum_{ij} w_i w_j },
\label{IQ_def}
\end{equation}
where $i$ and $j$ are pixel indices
and $w$ are the weights.
To avoid possible effects of $1/f$ noise,
we force the temperature map 
to come from a different frequency band than the polarization maps,
and thus use the temperature map at 61 GHz (V band)
for all correlations
except the V-band polarization maps,
which we correlate against the 41 GHz (Q band) temperature map.
Since {\sl WMAP} has a high signal-to-noise ratio measurement
of the CMB temperature anisotropy,
we use unit weight ($w_i = 1$) for the temperature maps
and noise weight ($w_j = N_j / \sigma_0^2 $)
for the polarization maps,
where $N_j$ is the effective number of observations in each pixel $j$
and $\sigma_0$ is the standard deviation of the white noise
in the time-ordered data
(Table 1 of \citet{bennett/etal:2003b}).
We compare the correlation functions
to Monte Carlo simulations of a null model,
which simulates the temperature anisotropy
using the best-fit $\Lambda$CDM model
\citep{spergel/etal:2003}
but forces the polarization signal to zero.
Each realization generates
a CMB sky in Stokes $I$, $Q$, and $U$ parameters,
convolves this simulated sky with the beam pattern 
for each differencing assembly,
then adds uncorrelated instrument noise to each pixel in each map.
We then co-add the simulated skies in each frequency band
and compute $C^{IQ}(\theta)$ using the same software
for both the {\sl WMAP} data
and the simulations.
All analysis uses only pixels outside the 
{\sl WMAP} Kp0 foreground emission mask
\citep{bennett/etal:2003c},
approximately 76\% of the full sky.

\begin{figure}
\plotone{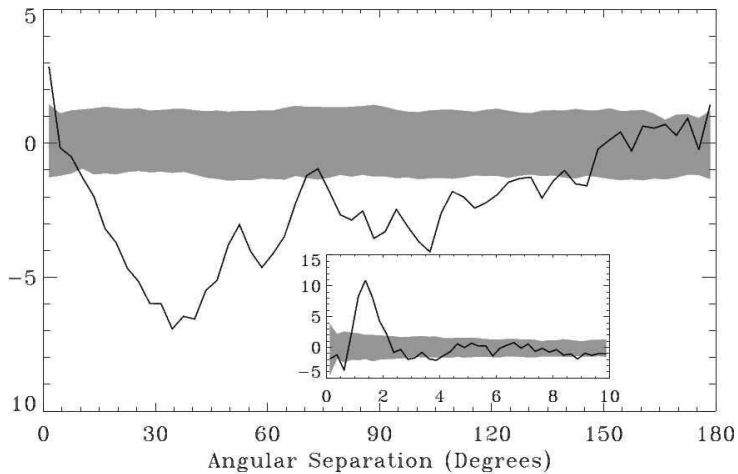}
\caption{Temperature-polarization correlation function
for {\sl WMAP} co-added QVW data.
The gray band shows the 68\% confidence interval for
similar co-added data
taken from Monte Carlo simulations without polarization.
The inset shows data for $\theta < 10\arcdeg$.
The data are inconsistent with no temperature-polarization
cross-correlations at more than 10 standard deviations.
Note that the data are not independent between angular bins.} 
\label{iq_corr_qvw_fig}
\end{figure}

Figure \ref{iq_corr_qvw_fig}
shows $C^{IQ}(\theta)$ 
derived by co-adding the individual
correlation functions
for the frequencies 41, 61, and 94 GHz
(Q, V, and W bands) 
least likely 
to be affected by Galactic foregrounds.
The grey band shows the 68\% confidence interval
for the null simulations.
It is clear that {\sl WMAP} detects 
a temperature-polarization signal
at high statistical confidence,
and that signals exist on both large and small angular scales.
We define a goodness-of-fit statistic
\begin{equation}
\chi^2 = \sum_{ab} 
~ [ C^{IQ}_{\rm MAP} - \langle C^{IQ}_{\rm sim} \rangle ]_a
~ {\bf M}^{-1}_{ab}
~ [ C^{IQ}_{\rm MAP} - \langle C^{IQ}_{\rm sim} \rangle ]_b ,
\label{chi_def}
\end{equation}
where
$C^{IQ}_{\rm MAP}$
is the co-added correlation function from {\sl WMAP} data,
$\langle C^{IQ}_{\rm sim} \rangle$
is the mean from the Monte Carlo simulations,
and 
${\bf M}$
is the covariance matrix 
between angular bins $a$ and $b$
derived from the simulations.
We find $\chi^2 = 207$ for 78 degrees of freedom
when comparing {\sl WMAP} to the null model:
{\sl WMAP} detects temperature-polarization correlations
significant at more than 10 standard deviations.

\subsection{Systematic Error Analysis}

Having detected a significant signal in the data,
we must determine whether this signal 
has a cosmological origin
or results from systematic errors or foreground sources.
We test the convergence of the mapping algorithm
using end-to-end simulations,
comparing maps derived from simulated time-ordered data 
to the input maps used to generate the simulated time series.
The simulations include all major instrumental effects,
including
beam ellipticity,
radiometer performance,
and instrument noise (including $1/f$ component),
and are processed using the same map-making software
as the {\sl WMAP} data
\citep{hinshaw/etal:2003b}.
The $Q$ and $U$ maps converge rapidly,
within the 30 iterations required to derive the calibration solution.
Correlations in the time-ordered data
introduce an anti-correlation in the $U$ map
at angles corresponding to the beam separation,
with amplitude 0.5\% of the noise in the map.
This effect is independent for each radiometer
and does not affect temperature-polarization cross-correlations.
Similarly, residual $1/f$ noise in the time series
can create faint striping in the maps,
but does not affect cross-correlations.

The largest potential systematic error 
in the temperature-polarization cross-correlation
results from 
bandpass mismatches in the amplification/detection chains.
We calibrate the {\sl WMAP} data
in thermodynamic temperature
using the Doppler dipole 
from the satellite's orbit about the Sun
as a beam-filling calibration source
\citep{hinshaw/etal:2003b}.
Astrophysical sources with a spectrum
other than a 2.7 K blackbody
are thus slightly mis-calibrated.
The amplitude is dependent on the product
of the source spectrum
with the unique bandpass of each radiometer.
If the bandpasses in each radiometer were identical,
the effect would cancel for any frequency spectrum,
but differences in the bandpasses 
between the two radiometers in each DA
generate a non-zero residual
in the difference signal used to generate polarization maps
(Eq. \ref{diff_def}).
This signal is spatially correlated 
with the unpolarized foreground intensity
but is independent of 
the orientation of the radiometers on the sky
(polarization angle $\gamma$).
In the limit of uniform sampling of $\gamma$
this term drops out of the sky map solution.
However, the {\sl WMAP} scan pattern does not 
view each pixel in all orientations;
unpolarized emission with a non-CMB spectrum
can thus be aliased into polarization
if the bandpasses of the two radiometers in each DA
are not identical.
This is a significant problem
only at 23 GHz (K band),
where the foregrounds are brightest
and the bandpass mismatch is largest.

\begin{figure}
\epsscale{0.6}
\plotone{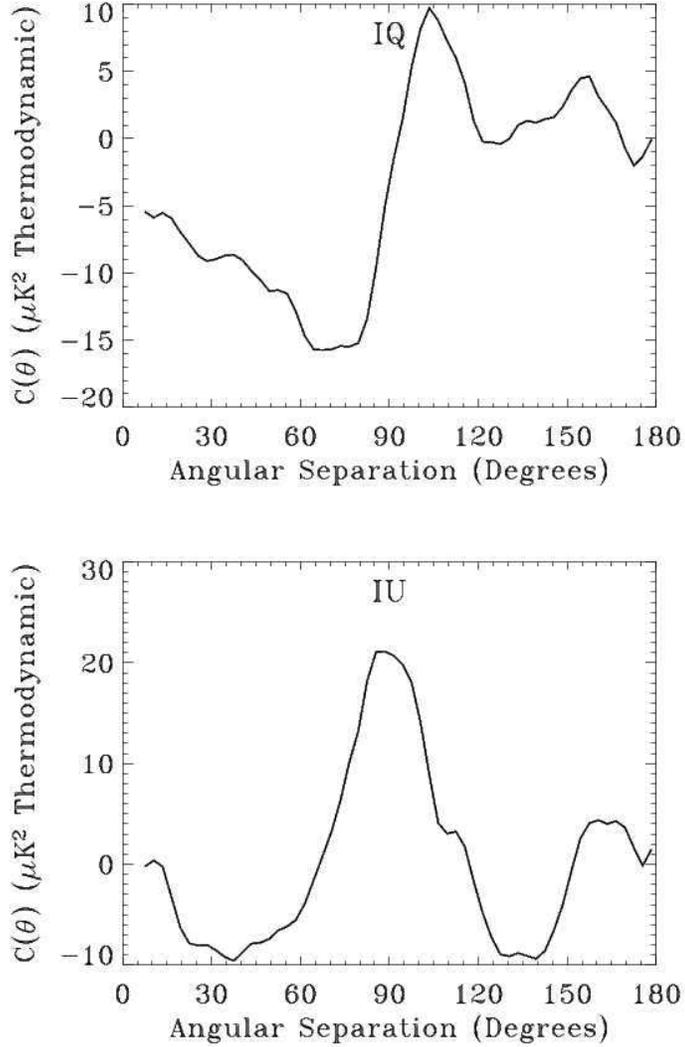}
\caption{Angular templates for
potential systematic errors
caused by bandpass mismatch 
between the 2 radiometers in each differencing assembly.
We fit this template to the correlation functions
from each DA
to detect or limit systematic errors
related to bandpass mismatch in the main beam.
The effect is significant only in K  and Ka bands,
which have the brightest unpolarized foregrounds.}
\label{bandpass_syserr}
\end{figure}

We quantify the effect of bandpass mismatch
using end-to-end simulations.
For each time-ordered sample, 
we compute the signal in each radiometer
using an unpolarized foreground model
and the measured pass bands in each output channel
\citep{jarosik/etal:2003}.
We then generate maps from the simulated data
using the {\sl WMAP} one-year sky coverage
and compute $C^{IQ}(\theta)$
using the output $I$, $Q$, and $U$ maps from the simulation.
Figure \ref{bandpass_syserr}
shows the predicted signal at K band.
We treat this as an angular template
and compute the least-squares fit of the
{\sl WMAP} data to this bandpass template
to determine the amplitude of the effect 
in the observed correlation functions.
We correct the {\sl WMAP} correlation functions 
$C^{IQ}(\theta)$ and $C^{IU}(\theta)$
at K and Ka bands 
by subtracting the best-fit template amplitudes.
The fitted signal has peak amplitude 
of 8 $\mu$K$^2$ at 23 GHz
and 5 $\mu$K$^2$ at 33 GHz.
No other channel has a statistically significant detection of this effect.

Sidelobe pickup of polarized emission from the Galactic plane
can also produce spurious polarization at high latitudes
in the $Q$ and $U$ maps.
We estimate this effect
using the measured far-sidelobe response
for each beam in each polarization
\citep{barnes/etal:2003}.
The simplest approach would be
to estimate the signal in each time-ordered sample,
convolving the full sky sidelobe response
with the Stokes $I$, $Q$, and $U$ maps
given the instantaneous orientation of the beams for each sample.
Such an approach is computationally expensive.
We instead approximate the signal in each pixel
by convolving the full sky sidelobe response
with the one-year $Q$ and $U$ maps.
For each pixel, we fix one beam on that pixel
while sweeping the other beam through
all orientations achieved in flight.
The average from the convolution
yields the sidelobe contribution for the pixel in question.
Details of the sidelobe maps
are presented in 
\citet{barnes/etal:2003}.
We correlate the sidelobe maps 
with the temperature anisotropy maps in each channel
to estimate the systematic error 
in the temperature-polarization correlation.
Sidelobe pickup of polarized structure in the Galactic plane
is less than 1 $\mu$K$^2$ in $C^{IQ}(\theta)$ at 23 GHz
and below 0.1 $\mu$K$^2$ in all other bands.
The effect of bandpass mismatch in the far sidelobes
(as opposed to the main beam)
is similarly weak,
with limits 1.3 $\mu$K$^2$ at 23 GHz
and less than 0.05 $\mu$K$^2$ in all other bands.
We correct the polarization maps for the estimated sidelobe signal
and propagate the associated systematic uncertainty
throughout our analysis.
Note that all of these systematic errors
depend on the Galactic foregrounds,
and have different frequency dependence than CMB polarization. 

Other instrumental effects are negligible.
We measure polarization 
by differencing the outputs of the two radiometers
in each differencing assembly
(Eq. \ref{diff_def}).
Calibration errors
(as opposed to the bandpass effect discussed above)
can alias temperature anisotropy
into a spurious polarization signal.
We have simulated the 
uncertainty in the calibration solution
using both realistic gain drifts
and drifts ten times larger than observed in flight
\citep{hinshaw/etal:2003b}.
Gain drifts
(either intrinsic or thermally-induced)
contribute less than 1 $\mu$K$^2$ to $C^{IQ}(\theta)$ 
in the worst band.

Null tests provide an additional check for systematic errors.
Thomson scattering of scalar temperature anisotropy
produces a curl-free polarization pattern.
A non-zero cosmological signal is thus expected 
only for the IQ (TE) correlation,
whereas systematic errors or foreground sources 
can affect both the IQ and IU (TB) correlations.
We also test linear combinations of radiometer maps
which cancel the polarization signal
but which test for systematic effects.
We compute the IQ and IU correlation functions
by correlating 
the Stokes $I$ sum map from the Q- or V-band (as noted above)
with the polarization {\it difference} maps
$(Q1 - Q2)/2$, 
$(V1 - V2)/2$, 
$(W1 - W2)/2$, 
and
$(W3 - W4)/2$.
We then co-add the results with their noise weights,
and compare the co-added result
for the polarization difference maps
to a similar computation for the polarization sum maps.
The temperature (Stokes $I$) map in all cases is a sum map;
the test is thus primarily sensitive to systematic errors
in the polarization data.

Table \ref{sum_diff_chi_table}
shows results of the null tests.
We compare $C^{IQ}(\theta)$ and $C^{IU}(\theta)$
for the sum and difference maps
to a null hypothesis
that the data consist of Stokes $I$ and instrument noise,
with no polarization in the Stokes $Q$ or $U$ maps.
We break the data into 2 angular regimes
to differentiate between signals at decoupling vs reionization.
We find a clear signal detection for $C^{IQ}(\theta)$ in the sum map
for both angular scales.
All other tests are consistent with instrument noise --
there is no evidence for additional systematic errors
in the temperature-polarization cross-correlation.

\subsection{Foregrounds}

Galactic emission is not a strong contaminant
for CMB temperature anisotropy,
but could be significant in polarization.
{\sl WMAP} measurements of unpolarized foreground emission
show synchrotron,
free-free,
and thermal dust emission
all sharing significant spatial structure
\citep{bennett/etal:2003c}.
Of these components,
only synchrotron emission
is expected to generate significant polarization;
other sources such as spinning dust
are limited to less than 5\% of the total intensity at 33 GHz.

Synchrotron emission from electrons accelerated in the Galactic magnetic field 
is the dominant unpolarized foreground 
at frequencies below $\sim$50 GHz.
Although it is known to be linearly polarized,
previous radio surveys
provide little guidance for the high-latitude polarization
at mm wavelengths.
Extrapolation of radio polarization maps
\citep{brouw/spoelstra:1976}
to millimeter wavelengths
indicate a polarization fraction between 10\% and 50\%
depending on Galactic latitude
\citep{lubin/smoot:1981}.
The unpolarized component
has angular power spectrum $c_\ell \propto \ell^{-2}$,
while the CMB power spectrum
rises to a set of peaks on angular scales $\theta \approx 1\arcdeg$
({\it cf} Fig 10(b) of \citet{bennett/etal:2003c}).
The angular dependence of the {\it polarized} foreground component
is expected to be even steeper
\citep{baccigalupi/etal:2001,
bruscoli/etal:2002,
tucci/etal:2002},
suggesting that foreground polarization
is most likely to affect temperature-polarization
correlations on large angular scales.
Radio maps at low Galactic latitude,
however,
demonstrate that the polarization intensity
is not necessarily well correlated with the unpolarized intensity,
complicating template analysis for 
temperature-polarization cross-correlations
\citep{uyaniker/etal:1998, 
uyaniker/etal:1999}.
We thus use the frequency dependence of the
measured temperature-polarization cross-correlation
to separate cosmic from foreground signals.

Foreground polarization above 40 GHz is faint:
fitting the correlation functions
at 41, 61, and 94 GHz
(Q, V, and W bands)
to a single power-law
$C^{IQ}(\theta,\nu) = C_0^{IQ}(\theta) ~  (\nu / \nu_0)^{\beta}$
yields spectral index $\beta = -0.4 \pm 0.4$,
consistent with a CMB signal ($\beta = 0$) 
and inconsistent with
the spectral indices expected for
synchrotron ($\beta \approx -3$),
spinning dust ($\beta \approx -2$),
or thermal dust ($\beta \approx 2$).
The measured signal can not be produced solely by 
a single foreground emission component
(unless the fractional polarization of the foreground emission
has a compensating frequency dependence,
which seems unlikely).

\begin{figure}
\epsscale{1.0}
\plotone{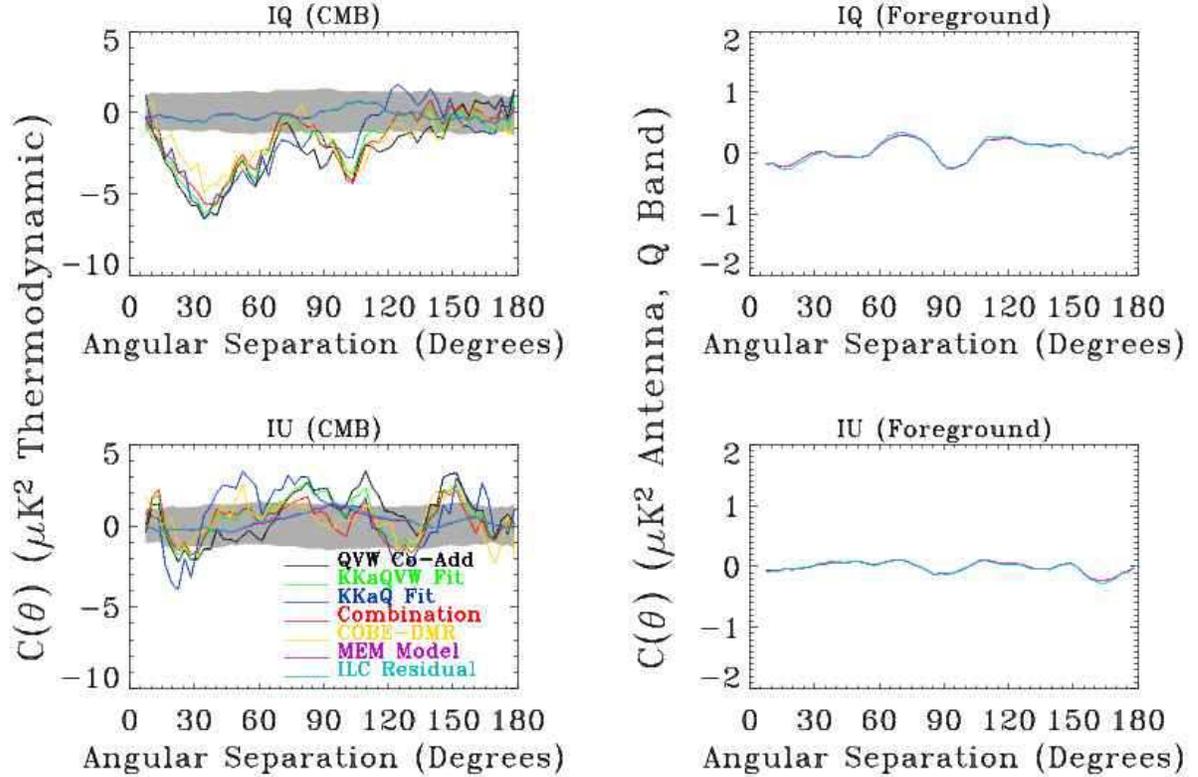}
\caption{Fitted CMB (left) and foreground (right) components
from a multi-frequency decomposition
of the measured two-point correlation functions.
Top panels show the IQ (TE) correlation,
while bottom panels show IU (TB).
The CMB component is shown in units of thermodynamic temperature,
while the foreground is shown in antenna temperature
evaluated at 41 GHz.
Different colors show the effect of
using different temperature maps
in the cross-correlation,
or including different polarization frequency channels
in the CMB-foreground decomposition.
``Co-Add'' refers to a noise-weighted linear combination
of the correlation functions computed for individual frequency channels.
``Fit'' refers to a 2-component fit (Eq. \ref{gal_fit_eqn})
using the specified polarization frequency channels.
The grey band shows the 68\% confidence interval
for the CMB component for the KKaQVW fit
(which has the smallest statistical uncertainty)
assuming CMB temperature anisotropy and instrument noise,
but no CMB polarization.
``Combination'' and ``COBE-DMR'' replace the
temperature map in Eq. \ref{IQ_def}
with maps with reduced foreground emission:
either the {\sl WMAP} internal linear combination map
or the COBE-DMR map of the CMB temperature.
``MEM Model'' and ``ILC Residual'' replace the
temperature map in Eq. \ref{IQ_def}
with maps dominated by foreground emission:
either the {\sl WMAP} maximum-entropy foreground model
or the residual map produced by subtracting
the internal linear combination map
from the individual temperature maps at each frequency.
The fitted CMB component
is stable as different frequency channels and data sets are analyzed.
Foreground emission is faint compared to the cosmic signal.
}
\label{cmb_gal_fig}
\end{figure}

A two-component fit
\begin{equation}
C^{IQ}(\theta,\nu) = C_{\rm CMB}^{IQ}(\theta) ~ 
+ ~ C_{\rm Gal}^{IQ}(\theta)  
\left( \frac{\nu}{\nu_0} \right)^{\beta}
\label{gal_fit_eqn}
\end{equation}
tests for the superposition 
of a CMB component with a single foreground component.
Figure \ref{cmb_gal_fig} shows the 
resulting decomposition into CMB and foreground components.
We obtain a marginal detection of foreground component
with best-fit spectral index
$\beta = -3.7 \pm 0.8$
consistent with synchrotron emission.
We test for consistency
or possible residual systematic errors
by repeating the fit
using different temperature maps
and different combinations of {\sl WMAP} polarization channels.
The fitted CMB component 
(left panels of Fig. \ref{cmb_gal_fig})
is robust against all combinations of frequency channels
and fitting techniques.
Note the agreement in Fig. \ref{cmb_gal_fig}
between nearly independent data sets:
the co-added QVW data (uncorrected for foreground emission)
and the KKaQ data (corrected for foreground emission).
We obtain additional confirmation
by replacing the V-band temperature map 
in the cross-correlation (Eq. \ref{IQ_def})
with the ``internal linear combination'' temperature map
designed to suppress foreground emission
\citep{bennett/etal:2003c}.
The fitted CMB component does not change.
We test for systematic errors
by replacing the temperature map
with the COBE-DMR map of the CMB temperature
\citep{bennett/etal:1996},
excluding {\it any} instrumental correlation
between the temperature and polarization data.
Again, the results are unchanged.

We further constrain foreground contributions
by computing the cross-correlation
between the {\sl WMAP} polarization data
and temperature maps dominated by foregrounds.
We replace the temperature map in Eq. \ref{IQ_def}
with either the {\sl WMAP} maximum-entropy foreground model
\citep{bennett/etal:2003c}
or a ``residual'' foreground map 
created by subtracting 
the internal linear combination CMB map
from the individual {\sl WMAP} temperature maps.
We then correlate the foreground temperature map
against the {\sl WMAP} polarization data in each frequency band,
and fit the resulting correlation functions
to CMB and foreground components
(Eq. \ref{gal_fit_eqn}).
The two foreground maps provide nearly identical results.
The fitted CMB component has nearly zero amplitude,
consistent with the instrument noise.
The fitted foreground has amplitude 
$0.5 \pm 0.1 ~\mu$K$^2$
at $\nu_0$ = 41 GHz,
with best-fit index $\beta = -3.4$
consistent with synchrotron emission.

\section{POLARIZATION CROSS-POWER SPECTRA}

In a second analysis method,
we compute the angular power spectrum of the temperature-polarization
correlations using a quadratic estimator
(Appendix A).
The power spectrum is the Legendre transform 
of the two-point correlation function,
and is more commonly encountered for theoretical predictions.
We compute $c_l^{TE}$ and $c_l^{TB}$ 
individually for the each {\sl WMAP} frequency band,
using uniform weight for the temperature map
and noise weight for the polarization maps.
We then combine the angular power spectra,
using noise-weighted QVW data
for $l > 21$ where foregrounds are insignificant,
and a fit to CMB plus foregrounds
using all 5 frequency bands
for $l \le 21$.
Since foreground contamination is weak, 
we gain additional sensitivity in this analysis
by using the Kp2 sky cut
retaining 85\% of the sky.

\begin{figure}
\epsscale{1.0}
\plotone{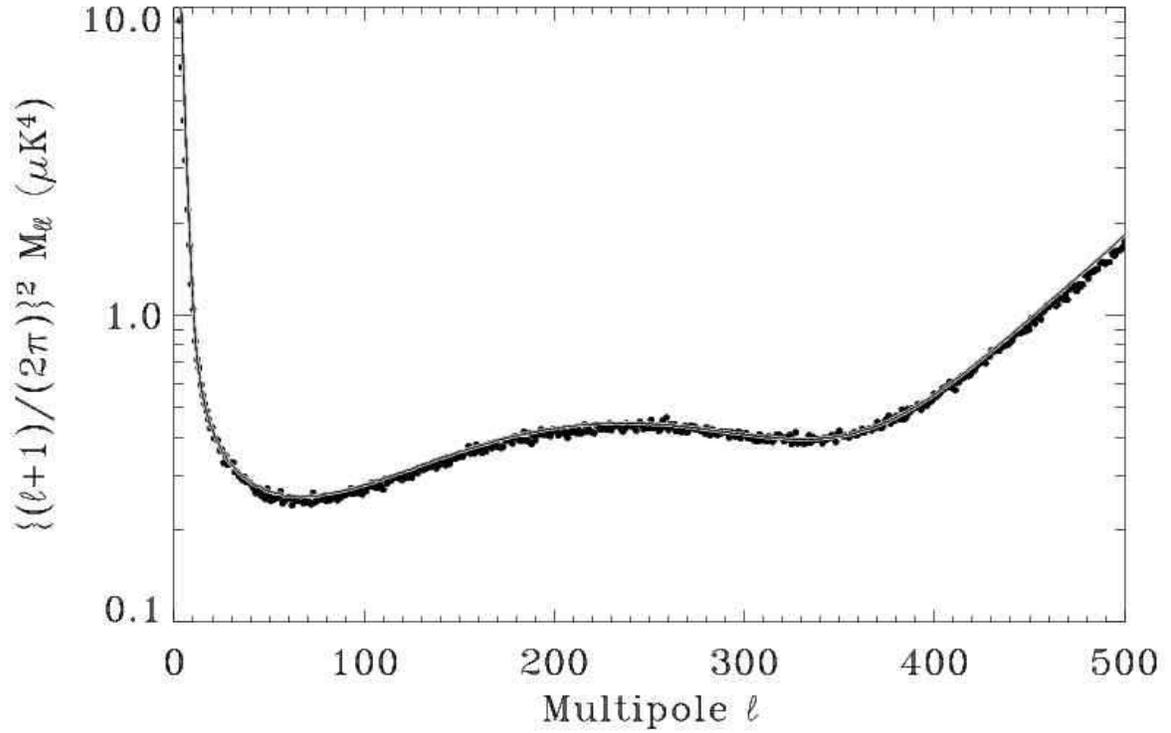}
\caption{Diagonal elements of the covariance matrix 
for the $c_l^{TE}$ polarization cross-power spectrum.  
Points show the diagonal elements computed from 
7500 Monte Carlo simulations.  
The solid line shows the analytical model
(Eq. \ref{covar_diag}). 
Note we multiply ${\bf M}_{ll}$ by 
$\left( \frac{l+1}{2 \pi} \right)^2$
to match the units in Figures
\ref{te_adiabatic_fig}
and
\ref{te_data_vs_model}.
}
\label{covar_diag_fig}
\end{figure}

\begin{figure}
\epsscale{1.0}
\plotone{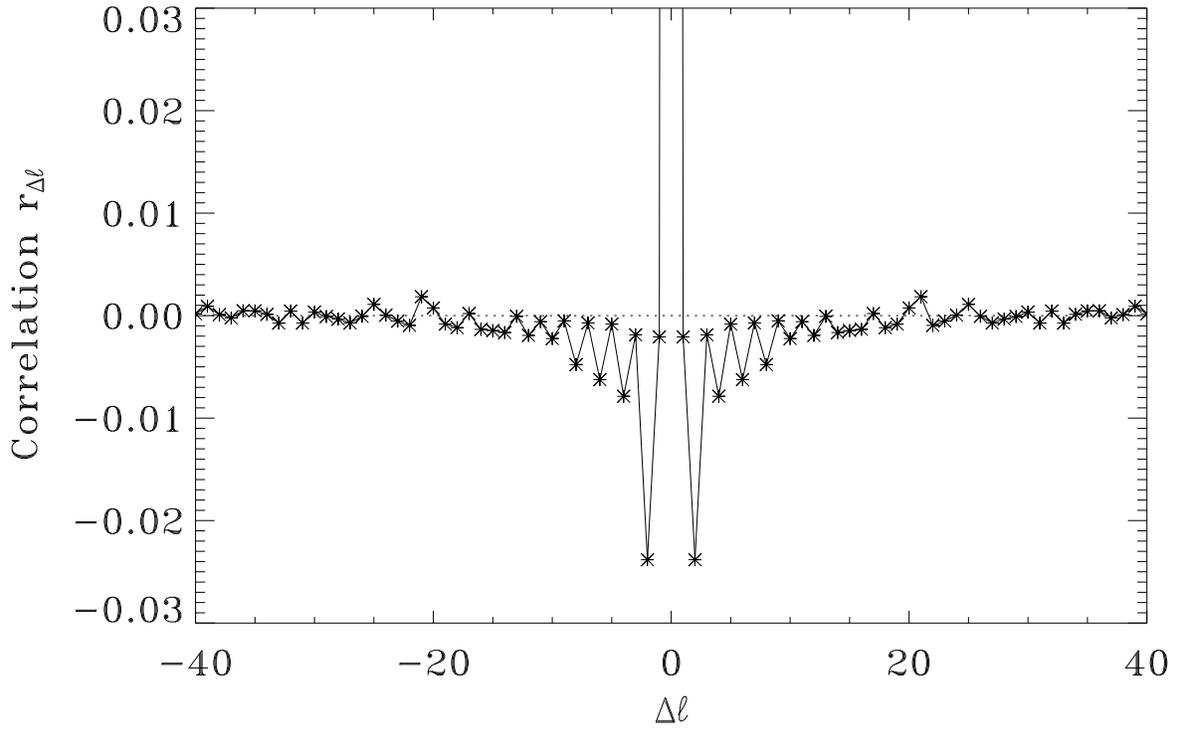}
\caption{Off-diagonal correlations $r_{\Delta l}$ in the covariance matrix
for the $c_l^{TE}$ polarization cross-power spectrum,
computed from simulations.
All values are normalized to $r_{\Delta l} = 1$ at $\Delta l = 0$.
The dotted line shows $r_{\Delta l} = 0$ for comparison.
The anti-correlation at $\Delta l = 2$
results from the spatial symmetry of the sky cut
and noise coverage.  }
\label{covar_off_diag_fig}
\end{figure}

We estimate the uncertainty in each $l$ bin
using the covariance matrix ${\bf M}$ 
for the polarization cross-power spectrum.
Based on our analysis of the $c_l^{TT}$ covariance matrix
\citep{hinshaw/etal:2003},
the $c_l^{TE}$ covariance matrix 
has the form along the diagonal of
\begin{eqnarray}
{\bf M}_{ll} &=&  <c_l^{TE} c_l^{TE}> - <c_l^{TE}>^2            \\
&&\simeq \frac{(c_l^{TT} +n_{TT}/w_l) (c_l^{EE}
+n_{EE}/w_l) + (c_l^{TE})^2}
{(2l+1) f_{sky} f_{sky}^{\rm eff}}
\label{covar_diag}
\end{eqnarray}
where
$n_{TT}$ and $n_{EE}$ are the TT and EE noise bias terms,
$w_l$ is the effective window function for the combined maps
\citep{page/etal:2003b},
$c_l^{TT}$ and $c_l^{EE}$ are the 
temperature and polarization angular power spectra,
$f_{sky}=0.85$ is the fractional sky coverage for the Kp2 mask,
and 
$f_{sky}^{\rm eff} = f_{sky}/1.14$ for noise weighting.
We take the $c_l^{TT}$ term from the measured temperature power spectra
\citep{hinshaw/etal:2003}
and the $c_l^{EE}$ term 
predicted by the best-fit $\Lambda$CDM model
\citep{spergel/etal:2003}
(allowing $c_l^{EE}$ to vary as a function of optical depth
in the likelihood analysis).
Figure \ref{covar_diag_fig}
compares the analytic expression for the diagonal elements
of the covariance matrix
to the mean derived from 7500 Monte Carlo simulations.
The analytic form (Eq. \ref{covar_diag})
accurately describes the simulations.
We approximate the off-diagonal terms
using the geometric mean of the covariance matrix terms
for uniform and noise weighting
\citep{hinshaw/etal:2003},\footnote{
Note that \citet{hinshaw/etal:2003} 
define off-diagonal elements
in terms of the inverse covariance matrix,
which differs from $r_{\Delta l}$ by a sign.}
\begin{equation}
{\bf M}_{ll'} \simeq  (~ {\bf M}_{ll} {\bf M}_{l'l'} ~)^{0.5} ~r_{\Delta l} .
\label{covar_off_diag}
\end{equation}
Figure \ref{covar_off_diag_fig}
shows the off-diagonal terms $r_{\Delta l}$
measured from Monte Carlo simulations.
The largest contribution, $-2.8\%$, 
is at $\Delta l = 2$ 
from the symmetry of our sky cut and noise coverage.  
The total anticorrelation is
$\sum_{\Delta l \ne 0} r_{\Delta l} = -0.124$.  
Because of this anti-correlation, 
the error bars for the binned $c_l^{TE}$ 
are slightly smaller than the naive estimate.
A second method of estimating the errors relies on 
end-to-end simulations
derived from simulated time-ordered data
consisting solely of instrument noise
(including the estimated contribution from $1/f$ fluctuations).
We have generated 11 ``noise'' 
sky maps each in Stokes $I$, $Q$, and $U$
and compute the variance
in TE directly from the variance in the simulated signal.  
These two approaches yield errors that are consistent 
to better than 5\%.
Since there are $2 l + 1$ multipoles at each $l$ value,
the fractional uncertainty expected in the Monte Carlo variance is
$[ 2 / (11 (2l+1) f_{sky}) ]^{0.5}$,
in agreement with this result.

\begin{figure}
\epsscale{1.0}
\plotone{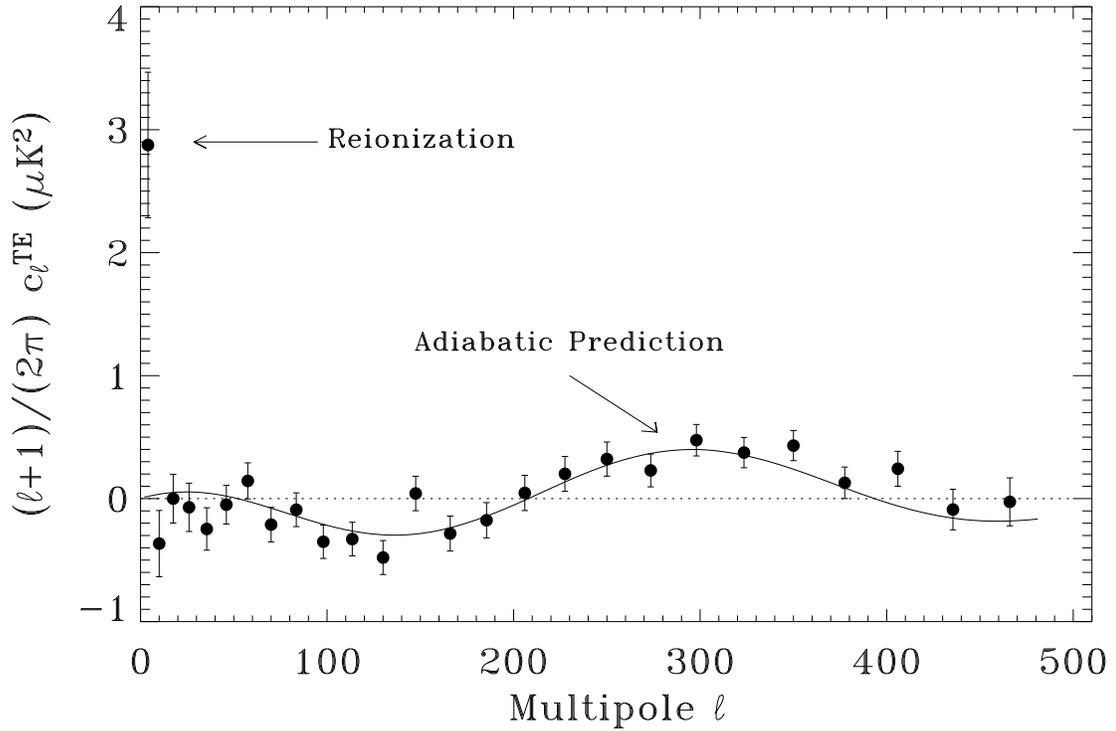}
\caption{Polarization cross-power spectra $c_\ell^{TE}$ 
for the {\sl WMAP} one-year data.
Note that we plot 
$(l+1)/2\pi~ c_l^{TE}$
and not
$l(l+1)/2\pi~ c_l^{TE}$.
This choice emphasizes the oscillatory nature of $c_\ell^{TE}$.
For clarity, the dotted line shows $c_l = 0$.
The solid line is the predicted signal
based on the $c_\ell^{TT}$ power spectrum of temperature anisotropy --
there are no free parameters.
The TE correlation on degree angular scales
($l > 20$)
is in excellent agreement
with the signal expected from adiabatic CMB perturbations.
The excess power at low $l$ indicates significant reionization
at large angular scales.}
\label{te_adiabatic_fig}
\end{figure}

Figure \ref{te_adiabatic_fig}
shows the polarization cross-power spectra 
for the {\sl WMAP} one-year data.
The solid line shows the predicted signal
for adiabatic CMB perturbations,
based only on a fit
to the measured temperature angular power spectrum $c_l^{TT}$ 
\citep{spergel/etal:2003,
hinshaw/etal:2003}.
Two features are apparent.
The TE data on degree angular scales
($l > 20$)
are in excellent agreement with
{\it a priori} predictions
of adiabatic models
\citep{coulson/crittenden/turok:1994}.
Other than the specification of adiabatic perturbations,
there are no free parameters --
the solid line is not a fit to $c_l^{TE}$.
The $\chi^2$ of 24.2 for 23 degrees of freedom
indicates that the CMB anisotropy
is dominated by adiabatic perturbations.
On large angular scales ($l < 20$)
the data show excess power compared to adiabatic models,
suggesting significant reionization.

The {\sl WMAP} detection of the acoustic structure 
in the TE spectrum
confirms several basic elements of the standard paradigm.  
The amplitudes of the peak and anti-peak
are a measure of the thickness 
of the decoupling surface,
while the shape confirms the assumption
that the primordial fluctuations are adiabatic.
Adiabatic fluctuations predict a temperature/polarization signal 
anticorrelated on large scales,
with TE peaks and anti-peaks located 
midway between the temperature peaks
\citet{hu/sugiyama:1994}.
The existence of TE correlations on degree angular scales
also provides evidence for super-horizon temperature fluctuations
at decoupling, as expected for inflationary models of cosmology
\citep{peiris/etal:2003}

\section{TEMPLATE POWER SPECTRA}
\label{section:template}

Figure \ref{te_adiabatic_fig} demonstrates 
that the power spectrum of temperature-polarization correlations
on degree angular scales
can be predicted 
using the power spectrum of the temperature fluctuations alone.
We use this 
for a third derivation of the TE cross-power spectrum,
based on template matching in pixel space.
For pixel sizes of a few degrees,
the signal-to-noise ratio for the temperature maps
is much larger than one per multipole,
while the S/N ratio in the polarization maps
is much less than one.
The likelihood function for the polarization measurement
then has the simple form
\begin{equation}
\log {\cal L} = (\hat P - \sum_l \alpha_l\hat P_{\rm pred}^l)^T N^{-1}
        (\hat P -\sum_l \alpha_l \hat P_{\rm pred}^l) ,
\end{equation}
where $\hat P$ is the measured polarization signal 
(a 2 $N_{\rm pixel}$ vector),
$\alpha_l = c_l^{TE}/c_l^{TT}$ 
is the polarization fraction at each $l$, 
$N$ is the pixel noise correlation matrix 
(a $2 N_{\rm pixel} \times 2 N_{\rm pixel}$ matrix) 
and
\begin{eqnarray} 
Q_{pred}^l(\hat n) = \sum_m a_{lm}   (_{2}Y_{lm}(\hat n) + _{-2}Y_{lm}(\hat n)) \nonumber \\
U_{pred}^l(\hat n) = i\sum_m a_{lm}  (_{2}Y_{lm}(\hat n) - _{-2}Y_{lm}(\hat n)) .
\end{eqnarray}
Here $_{\pm 2}Y_{lm}(\hat n)$ are the spin harmonics,
while $a_{lm}$ are the measured coefficients 
for an all-sky map of the CMB temperature.
Imposing a cut to mask the Galactic plane
introduces additional correlations;
we avoid this by using the ``internal'' 
linear combination temperature map
\citep{bennett/etal:2003c}
without imposing a sky cut.

The maps
$Q_{pred}$ and $U_{pred}$
represent the predicted polarization pattern
based on the observed pattern of temperature anisotropy.
We fit these template maps
to the observed $Q$ and $U$ polarization maps
to derive the polarization fraction $\alpha_l$
and thus the $c_l^{TE}$ polarization cross-power spectrum.
Minimizing the likelihood function
yields the normal equations
\begin{equation}
K_{ll'} \alpha_{l'} = y_{l} ,
\end{equation}
where
\begin{equation}
y_l = \hat P  N^{-1} \hat P_{pred}^l
\end{equation}
and
\begin{equation}
K_{ll'} = \hat P_{pred}^l N^{-1} \hat P_{pred}^{l'} .
\end{equation}
These equations show the advantages of this approach.
We compare the data with a template in pixel space,
making it straightforward
to include a spatially varying noise signal.
We directly compare the measured polarization maps
to a prediction based on the measured temperature maps,
yielding a measurement of the TE polarization cross-power spectrum 
in the observed sky
unaffected by cosmic variance.
We can thus more easily compute the errors on the 
measured polarization fraction.
The input temperature map (Stokes $I$)
is already corrected for foreground emission
(much simpler in pixel space 
where the unpolarized foregrounds are more easily measured),
greatly reducing the foreground contribution
to the cross-power spectra.

We thus compute the temperature-polarization cross-correlation
using three disparate techniques:
the two-point angular correlation function,
a quadratic estimator for the power spectrum in Fourier space,
and a template fit in pixel space.
All methods are in good agreement
despite their very dissimilar treatment of the data.
All methods show a significant excess of power
for $l < 10$.

\section{REIONIZATION}

{\sl WMAP} detects statistically significant correlations 
between the CMB temperature and polarization.
The signal on degree angular scales ($l > 20$)
agrees with the signal expected in adiabatic models
based solely on the temperature power spectrum,
without any additional free parameters.
We also detect power on large angular scales ($l < 10$)
well in excess of the signal 
predicted by the temperature power spectrum alone.
This signal can not be explained by data processing,
systematic errors,
or foreground polarization,
and has a frequency spectrum consistent with a cosmological origin.

The signal on large angular scales
has a natural interpretation
as the signature of early reionization.\footnote{
Although tensor modes
can also generate TE correlations at large angular scales,
tensor-to-scalar ratios $r$
large enough to fit the {\sl WMAP} TE data
are ruled out by the {\sl WMAP} TT data
\citep{spergel/etal:2003}.}
Both the temperature and temperature-polarization power spectra
can be related to the 
power spectrum of the radiation field 
during scattering
\citep{zaldarriaga:1997}.
Thomson scattering
damps the temperature anisotropy
and regenerates a polarized signal
on scales comparable to the horizon.
The existence of polarization on scales
much larger than the acoustic horizon at decoupling
implies significant scattering at more recent epochs. 

\subsection{Reionization in a $\Lambda$CDM Universe}

If we assume that the $\Lambda$CDM model is the best description 
of the physics of the early universe, 
we can fit the observed
temperature-polarization cross-power spectrum
to derive the optical depth 
\ensuremath{\tau}.
We assume a step function for the ionization fraction $x_e$
and use the CMBFAST code 
\citep{seljak/zaldarriaga:1996}
to predict the multipole moments
as a function of optical depth. 
While this assumption is simplistic,
our conclusions on optical depth 
are not very sensitive to details of the reionization history
or the background cosmology.

\begin{figure}
\plotone{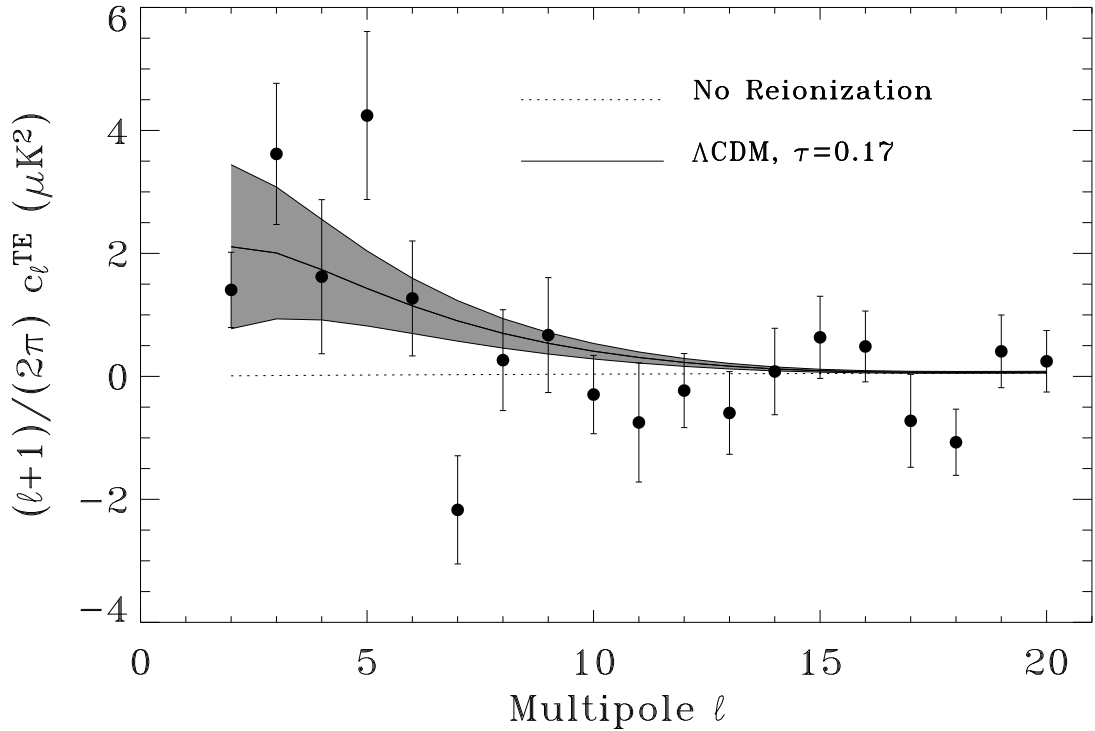}
\caption{{\sl WMAP} Polarization cross-power spectra $c_\ell^{TE}$
(filled circles)
compared to $\Lambda$CDM models 
with and without reionization.
The rise in power for $l < 10$
is consistent with reionization
optical depth \ensuremath{\tau = 0.17 \pm 0.04}.
The error bars on {\sl WMAP} data
reflect measurement errors only;
adjacent points are slightly anti-correlated.
The grey band shows the 68\% confidence interval 
from cosmic variance.
The value at $l=7$ is particularly sensitive
to the foreground correction.
}
\label{te_data_vs_model}
\end{figure}

\begin{figure}
\plotone{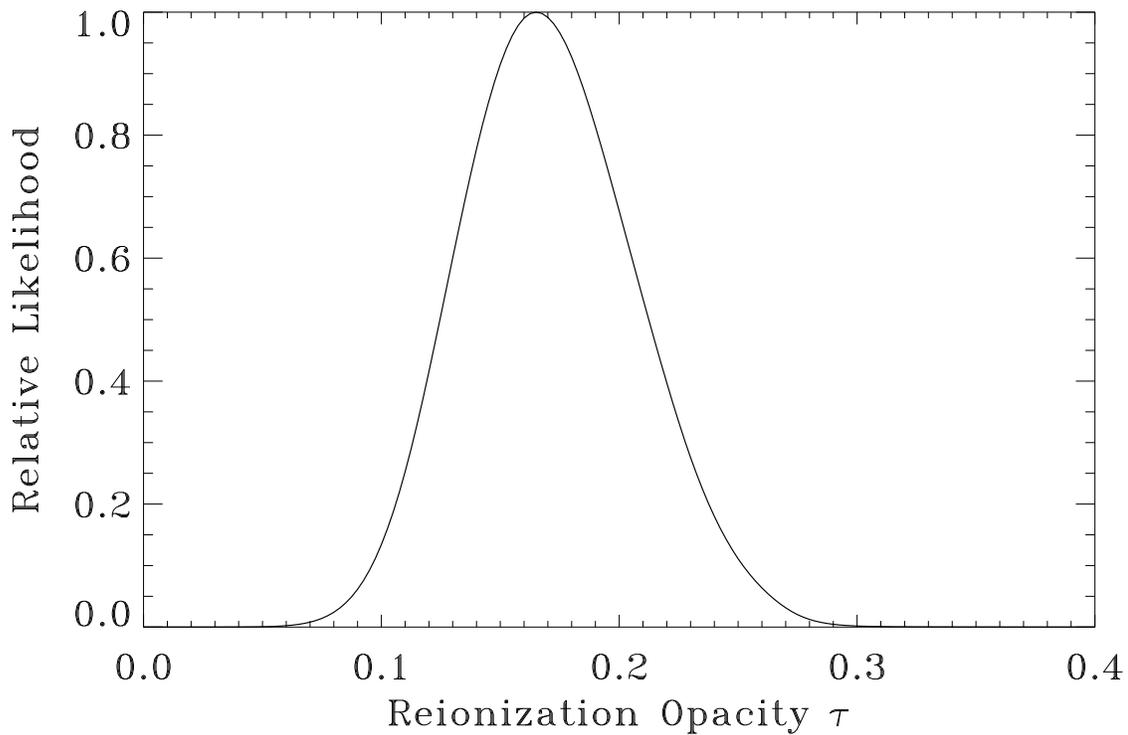}
\caption{Likelihood function for optical depth $\tau$
for a $\Lambda$CDM cosmology,
using all 5 {\sl WMAP} frequency bands
fitted to CMB plus foregrounds with foreground spectral index
$\beta = -3.7$.
After including systematic and foreground uncertainties
the optical depth is consistent with a value $\tau=0.17$
with 95\% confidence range $0.09 \le \tau \le  0.28$.
}
\label{tau_like_fig}
\end{figure}

Figure \ref{te_data_vs_model}
compares the polarization cross-power spectrum
$c_l^{TE}$
derived from the quadratic estimator
to $\Lambda$CDM models
with and without reionization.
The rise in power for $l < 10$
is clearly inconsistent with no reionization.
We quantify this using a maximum-likelihood analysis
\begin{equation}
{\cal L} \propto
\frac{ \exp( -\frac{1}{2} \chi^2 )}
     { |{\bf M}|^{1/2} } .
\label{like_def}
\end{equation}
Figure \ref{tau_like_fig} shows
the relative likelihood ${\cal L} / {\rm Max}( {\cal L})$
for the optical depth \ensuremath{\tau}
assuming a $\Lambda$CDM cosmology,
with all other parameters
fixed at the values derived 
from the temperature power spectrum alone
\citep{spergel/etal:2003}.
The likelihood 
for the 5-band data
corrected for foreground emission
peaks at $\tau = 0.17 \pm 0.03$ (statistical error only):
{\sl WMAP} detects the signal from reionization at high statistical confidence.

A full error analysis for \ensuremath{\tau}
must account for systematic errors and foreground uncertainties.
We propagate these effects
by repeating the maximum likelihood analysis
using different combinations of {\sl WMAP} frequency bands
and different systematic error corrections.
We correct $C^{IQ}(\theta)$ in each frequency band
not for the best estimate of the systematic error templates,
but rather the best estimate plus or minus one standard deviation.
We then fit the mis-corrected $C^{IQ}(\theta,\nu)$ for
a CMB piece plus a foreground piece (Eq. \ref{gal_fit_eqn})
and use the CMB piece in a maximum-likelihood analysis for 
\ensuremath{\tau}.
The change in the best-fit value for \ensuremath{\tau}
as we vary the systematic error corrections
propagates the uncertainties in these corrections.
Systematic errors have a negligible effect
on the fitted optical depth;
altering the systematic error corrections
changes the best-fit values of \ensuremath{\tau} 
by less than 0.01.

The largest non-random uncertainty is the foreground separation.
We assess the uncertainty in the foreground separation
by repeating the entire systematic error analysis
(using both standard and altered systematic error corrections)
with the foreground spectral index $\beta = -3.7 \pm 0.8$
shifted one standard deviation up or down from the best-fit value.
Table \ref{tau_table}
shows the fitted optical depth \ensuremath{\tau}
and goodness-of-fit statistic $\chi^2$
for different data combinations and foreground spectral indices
derived from the analysis 
of the two-point correlation function $C^{IQ}(\theta)$.
The first set of rows shows values derived by simply co-adding
the {\sl WMAP} frequency channels,
without any correction for foregrounds.
Data at 41, 61, and 94 GHz 
(Q, V, and W bands)
where foregrounds are negligible
show similar values for \ensuremath{\tau};
the $\chi^2 \approx 66$ for 57 degrees of freedom
indicates that the data
are in agreement with reionized models.
Adding additional low-frequency channels
reduces the formal statistical uncertainty,
but introduces non-zero foreground contamination
as shown by the marked increase in $\chi^2$.
The next three sets of rows 
show the results
when the data are separated 
into CMB and foreground components (Eq. \ref{gal_fit_eqn}).
All data combinations are now in agreement;
we obtain nearly identical values for \ensuremath{\tau}
when fitting either the highest-frequency data set QVW 
or the lowest-frequency set KKaQ.
The fitted optical depth is insensitive to the spectral index:
varying the spectral index from -2.9 to -4.5
changes the fitted values by 0.02 or less.
We adopt 
\ensuremath{\tau = 0.17 \pm 0.04}
as the best estimate for the optical depth to reionization,
where the error bar reflects
a 68\% confidence level interval
including statistical, systematic, and foreground uncertainties.

\citet{spergel/etal:2003} include the TE data
in a maximum-likelihood analysis
combining {\sl WMAP} data with other astronomical measurements.
The resulting value, 
\ensuremath{\tau = 0.17 \pm 0.06},
is consistent with 
the value derived from the TE data alone.
The larger uncertainty reflects
the effect of simultaneously 
fitting multiple parameters.
The TE analysis propagates foreground uncertainties
by re-evaluating the likelihood
using different foreground spectral index.
Since foreground affect only the lowest multipoles,
the combined analysis propagates foreground uncertainty
by doubling the statistical uncertainty in $c_l^{TE}$
for $2 \le l \le 4$ to account for this effect.

\subsection{Model-Independent Estimate}

An alternative approach avoids assuming any cosmological model and uses the
measured temperature angular correlation function to determine the radiation
power spectrum at recombination.  This approach assumes that the best estimate
of the three dimensional  radiation power spectrum 
is the {\it measured} angular power spectrum 
rather than a model fit to the angular power spectrum.
Given the observed temperature power spectrum
$c_l^{TT}$, 
we derive the predicted polarization cross-power spectrum $c_l^{TE}$
(\S\ref{section:template}),
which we then fit to the observed TE spectrum
as a function of optical depth \ensuremath{\tau}.
We obtain
$\tau = 0.16 \pm 0.04$,
in excellent agreement with
the value derived assuming a $\Lambda$CDM cosmology.
We emphasize that the model-independent technique
makes {\it no} assumptions
about the cosmology.
The fact that it agrees well with 
the best-fit model
from the combined temperature and polarization data
\citep{spergel/etal:2003}
is an additional indication that the
observed temperature-polarization correlations
on large angular scales
represent the imprint
of physical conditions at reionization.
The dependence on the underlying cosmology is small.

\subsection{Early Star Formation}

Reionization can also be expressed 
as a redshift \ensuremath{z_r}
assuming an ionization history.
We consider two simple cases.
For instantaneous reionization with
ionization fraction $x_e = 1$ 
at $z < \ensuremath{z_r}$,
the measured optical depth corresponds
to redshift $\ensuremath{z_r} = 17 \pm 3$.
This conflicts with measurements of the
Gunn-Peterson absorption trough in spectra of distant quasars,
which show neutral hydrogen present at $z \approx 6$
\citep{becker/etal:2001,
djorgovski/etal:2001,
fan/etal:2002}.
Reionization clearly did not occur through a single rapid phase transition.
However,
since absorption spectra are sensitive 
to even small amounts of neutral hydrogen,
models with partial ionization $x_e \lesssim 1$
can have enough neutral column density
to produce the Gunn-Peterson trough
while still providing free electrons
to scatter CMB photons and produce large-scale polarization.
Direct Gunn-Peterson observations 
only imply a neutral hydrogen fraction $\gtrsim$ 1\%
\citep{fan/etal:2002}.
Accordingly, we modify the simplest model
to add a second transition:
a jump from $x_e=0$ to $x_e=0.5$ at redshift
\ensuremath{z_r},
followed by a second transition
from $x_e=0.5$ to $x_e=1$ at redshift $z=7$.
Fitting this model
to the measured optical depth
yields 
$\ensuremath{z_r} \approx 20$.
In reality,
reionization is more complicated than simple step transitions.
Allowing for model uncertainty,
the measured optical depth
is consistent with reionization at redshift 
$11 < \ensuremath{z_r} < 30$,
corresponding to times 
$100 < \ensuremath{t_r} < 400$ 
Myr after the Big Bang 
(95\% confidence).

Extrapolations of
the observed ionizing flux 
to higher redshift lead to predicted
CMB optical depth between $0.04 - 0.08$
\citep{miralda-escude:2003},
lower than our best fit values.
The measured optical depth
thus implies additional sources of ionizing flux at high redshift.
An early generation of very massive (Pop III) stars
could provide the required additional heating.
\citet{tegmark/etal:1997} estimate that $10^{-3}$ of all baryons
should be in collapsed objects by $z=30$.
If these baryons form massive stars, 
they would reionize the universe.
However, photons below the hydrogen ionization threshold 
will destroy molecular hydrogen
(the principal vehicle for cooling in early stars),
driving the effective mass threshold 
for star formation to $\sim 10^8$ solar masses
and impeding subsequent star formation
\citep{haiman/rees/loeb:1997,
gnedin/ostriker:1997,
tegmark/etal:1997}.
X-ray heating and ionization
\citep{venkatesan/giroux/shull:2001,
oh:2001} 
may provide a loophole to this argument
by enhancing the formation of $H_2$ molecules
\citep{haiman/abel/rees:2000}.

\citet{cen:2003} provides a physically-motivated model 
of ``double reionization''
that resembles the two-step model above.
A first generation of massive Pop III stars
initially ionizes the intergalactic medium.
The increased metallicity of the intergalactic medium
then produces a transition to smaller Pop II stars,
after which the reduced ionizing flux allows 
regeneration of a neutral hydrogen fraction.
The ionization fraction remains at $x_e \approx 0.6$
until the global star formation rate
surpasses the recombination rate at $z=6$,
restoring $x_e=1$.
The predicted value
$ \ensuremath{\tau} = 0.10 \pm 0.03$
should be increased somewhat
to reflect the higher {\sl WMAP}
values for the baryon density $\Omega_b$ and normalization $\sigma_8$
\citep{spergel/etal:2003}.
The contribution from ionized helium
will also serve to increase $\tau$
\citep{venkatesan/tumlinson/shull:2003,
wyithe/loeb:2003}.
The {\sl WMAP} determination of the optical depth
indicates that ionization history
must be more complicated than a simple instantaneous step function.
While physically plausible models 
can reproduce the observed optical depth,
reionization remains a complex process
and can not be fully characterized by a single number.
A more complete determination of the ionization history
requires evaluation
of the detailed $TE$ and $EE$ power spectra
\citep{kaplinghat/etal:2003,
hu/holder:2003}.

\section{CONCLUSIONS}

{\sl WMAP} detects statistically significant correlations
between the temperature and polarization maps.
The correlations are inconsistent with instrument noise
and are significantly larger than the upper limits
established for potential systematic errors.
The correlations are present in all {\sl WMAP} frequency bands
with similar amplitude from 23 to 94 GHz;
fitting the data to a single power-law in frequency
yields a spectral index 
$\beta = -0.4 \pm 0.4$,
consistent with a CMB signal ($\beta = 0$) 
and inconsistent with
the measured spectral indices for
Galactic foreground emission.
A two-component fit to
a superposition of CMB and Galactic foregrounds
yields a positive foreground detection
in both curl- and curl-free modes,
with best-fit spectral index
$\beta = -3.7 \pm 0.8$
consistent with synchrotron emission
of amplitude $0.5 \pm 0.1 ~\mu{\rm K}^2$
antenna temperature at 41 GHz.

The fitted CMB component is robust
against different data combinations
and fitting techniques.
On small angular scales
$(\theta < 5\arcdeg)$, 
the {\sl WMAP} data 
show the temperature-polarization
expected from adiabatic perturbations in the temperature power spectrum.
The data for $\ell > 20$ agree well with the signal predicted
solely from the temperature power spectra,
with no additional free parameters.

The data show excess power on large angular scales
($\theta > 10\arcdeg$)
compared to the predictions based on the temperature power spectrum alone.
The excess power is well described by early reionization
at redshift 
$\ensuremath{z_r} = 20^{+10}_{-9}$,
corresponding to times 
$\ensuremath{t_r} = 180^{+220}_{-80}$ Myr after the Big Bang
(95\% confidence).
A model-independent fit to reionization optical depth
yields results consistent with the $\Lambda$CDM model.
Our best estimate for the optical depth is
\ensuremath{\tau = 0.17 \pm 0.04} (68\% confidence)
where the error terms include
statistical, systematic, and foreground uncertainties.
This value is larger than expected
given the detection of a Gunn-Peterson trough
in the absorption spectra of distant quasars,
and implies that 
the universe has a complex ionization history.

The {\sl WMAP} detection of early reionization
opens a new frontier
to explore the universe at redshift $6 < z < 30$.
{\sl WMAP}'s sensitivity to reionization
is currently limited by instrument noise,
both as direct statistical uncertainty
and in the ability to better model and remove faint polarized foregrounds.
Instrumental effects
do not limit analysis of temperature-polarization correlations.
The TE power spectrum and covariance matrix
are available at
{\tt http://lambda.gsfc.nasa.gov}.
We are currently performing a more complete set of 
systematic error analyses
in the individual $Q$ and $U$ maps.
A future data release will include 
full-sky polarization maps
and polarization power spectra.

\acknowledgements

The {\sl WMAP} mission is made possible by the support of the Office of Space 
Sciences at NASA Headquarters and by the hard and capable work of scores of 
scientists, engineers, technicians, machinists, data analysts, budget analysts, 
managers, administrative staff, and reviewers. 

\clearpage
\appendix

\section{Quadratic Estimator for Temperature-Polarization Power Spectrum}

We estimate the temperature-polarization power spectrum 
from pixelized sky maps using the following formalism.
We begin by expanding the temperature and polarization fluctuations in 
generalized spherical harmonics
\begin{equation}
T(\hat n) =\sum_{lm} a_{lm} Y_{lm}(\hat n) 
\label{harmonic_def_t}
\end{equation}
\begin{equation}
Q(\hat n) \pm i U(\hat n) =\sum_{lm} a_{\mp 2, lm}\ _{\mp 2}Y_{lm}(\hat n)  
\label{harmonic_def_qu}
\end{equation}
We then decompose the polarization fluctuations into $E$ and $B$ like pieces
\begin{equation}
a_{\pm 2, lm} = E_{lm} \pm i B_{lm} .
\label{harmonic_def_eb}
\end{equation} 
We can use the basic properties of the spherical harmonics 
\begin{eqnarray}
_NY_{lm}&=& (-1)^N \  _{-N}Y_{l\ -m}^* \\
\int d\hat n \ _NY_{lm}(\hat n) _NY^*_{l'm'}(\hat n) &=& \delta_{l}^{l'} \delta_{m}^{m'}
\end{eqnarray}
to derive
\begin{eqnarray}
E_{lm} = \frac{1}{2} \int d\hat n ~ & \left[ ~Q(\hat n) 
\left(\ _{2}Y^*_{lm}(\hat n) +\ _{-2}Y^*_{lm}(\hat n)\right) \right. \nonumber \\
&-i \left. U(\hat n)
\left(\ _{2}Y^*_{lm}(\hat n) -\ _{-2}Y^*_{lm}(\hat n)\right) ~ \right] \nonumber
\\
B_{lm} = -\frac{1}{2} \int d\hat n ~ & \left[ ~U(\hat n) 
\left(\ _{2}Y^*_{lm}(\hat n) +\ _{-2}Y^*_{lm}(\hat n)\right)\right.\nonumber \\
&+i \left. Q(\hat n)
\left(\ _{2}Y^*_{lm}(\hat n) -\ _{-2}Y^*_{lm}(\hat n)\right) ~ \right] .
\end{eqnarray}
We can now generalize the approach of
\citet{hivon/etal:2002}
to estimate the coupling terms.  
We multiply the temperature and polarization maps by
a weighting function
\begin{eqnarray}
\tilde T_{lm} &=& \int d\hat n ~w^T(\hat n) ~ T(\hat n) Y^*_{lm}(\hat n)  \\
\tilde E_{lm} &=& \frac{1}{2} \int d\hat n ~w^P(\hat n) ~\left[ Q(\hat n) 
\left(\ _{2}Y^*_{lm}(\hat n) +\ _{-2}Y^*_{lm}(\hat n)\right) \right.\nonumber \\
&&\left. -i U(\hat n)\
\left(\ _{2}Y^*_{lm}(\hat n) -\ _{-2}Y^*_{lm}(\hat n)\right) \right] \\
\tilde B_{lm} &=& -\frac{1}{2} \int d\hat n ~w^P(\hat n) ~\left[ U(\hat n) 
\left(\ _{2}Y^*_{lm}(\hat n) +\ _{-2}Y^*_{lm}(\hat n)\right) \right.\nonumber \\
&&\left. +i Q(\hat n) 
\left(\ _{2}Y^*_{lm}(\hat n) -\ _{-2}Y^*_{lm}(\hat n)\right) \right] .
\end{eqnarray}
We expand the weighting function in spherical harmonics
\begin{equation}
w(\hat n) = \sum_{lm} w_{lm} Y_{lm}(\hat n) ,  
\end{equation} 
and combine with equations 
\ref{harmonic_def_t} -- \ref{harmonic_def_eb}
to yield
\begin{eqnarray}
\tilde T_{lm} &=& \sum_{l'm'l''m''} w^T_{l'' m''} T_{l' m'}
\int d\hat n ~ Y_{l'm'}(\hat n) Y_{l'' m''}(\hat n)  Y^*_{lm}(\hat n) \nonumber \\
\tilde E_{lm} &=& \frac{1}{2} \sum_{l'm'l''m''} w^P_{l'' m''} \left[E_{l'm'}
\int d\hat n ~ Y_{l'' m''} (\hat n) 
\left(\ _{2}Y_{l'm'}(\hat n)\ _{2}Y^*_{lm}(\hat n)  
+\ _{-2}Y_{l'm'}(\hat n)\ _{-2}Y^*_{lm}(\hat n)\right) \right.\nonumber \\
&&\left. +i B_{lm}
\int d\hat n  ~ Y_{l'' m''} (\hat n) 
\left(\ _{2}Y_{l'm'}(\hat n)\ _{2}Y^*_{lm}(\hat n) 
-\ _{-2}Y_{l'm'}(\hat n)\ _{-2}Y^*_{lm}(\hat n)\right) \right]\nonumber \\
\tilde B_{lm} &=& \frac{1}{2} \sum_{l'm'l''m''} w^P_{l'' m''} \left[B_{l'm'}
\int d\hat n  ~ Y_{l'' m''} (\hat n) 
\left(\ _{2}Y_{l'm'}(\hat n)\ _{2}Y^*_{lm}(\hat n) 
+\ _{-2}Y_{l'm'}(\hat n)\ _{-2}Y^*_{lm}(\hat n)\right)\right.\nonumber \\
&&\left.-i E_{lm}
\int d\hat n  ~ Y_{l'' m''} (\hat n) 
\left(\ _{2}Y_{l'm'}(\hat n)\ _{2}Y^*_{lm}(\hat n) 
-\ _{-2}Y_{l'm'}(\hat n)\ _{-2}Y^*_{lm}(\hat n)\right) \right] .
\end{eqnarray}
We can then use
\begin{eqnarray}
\int d\hat n \ _N Y_{lm}^{*}(\hat n)\ _{N'} Y_{l'm'}(\hat n)
\ _{N''} Y_{l''m''} (\hat n) &=& (-1)^{N+m}\left[\frac{(2l+1)(2l'+1)(2l''+1)}
{4\pi}\right]^{1/2}\nonumber \\
&&\left( \begin{array}{ccc}
l & l' & l'' \\
-N & N' & N''
\end{array} \right)
\left( \begin{array}{ccc}
l & l' & l'' \\
-m & m' & m''
\end{array} \right)
\end{eqnarray}
to compute
\begin{equation}
\left( \begin{array} {c}
\tilde c_{l}^{TT} \\
\tilde c_{l}^{TE} \\
\tilde c_{l}^{TB} \\
\tilde c_{l}^{EE} \\
\tilde c_{l}^{BB} \\
\end{array} \right)
=
\left(\begin{array}{ccccc}
& & & & \\
& & & & \\
& &M_{ll'}^{ab} & & \\
& & & & \\
& & & & \\
\end{array}\right)
\left( \begin{array} {c}
c_{l'}^{TT} \\
c_{l'}^{TE} \\
c_{l'}^{TB} \\
c_{l'}^{EE} \\
c_{l'}^{BB} \\
\end{array} \right) .
\end{equation}
These expressions can be reduced using the symmetry and orthogonality
properties of 3-$j$ symbols, as given in Eqs. 1.8 and 1.14 of 
\citet{rotenberg/etal:1959}.  In particular, imaginary terms drop out,
and summations over products of 3-$j$ symbols with $-m$, $m'$ and $m''$ 
in the bottom row evaluate to $1/(2l''+1)$.  
After some algebra, the coupling terms reduce to
\begin{eqnarray}
M_{ll'}^{TT,TT}  &=& \frac{(2l'+1)}{4\pi}\sum_{l''} {\cal W}^{TT}_{l''} 
\left( \begin{array}{ccc}
l & l' & l'' \\
0& 0 &0 
\end{array} \right)^2\nonumber  \\
M_{ll'}^{TE,TE}  &=& M_{ll'}^{TB,TB} \\
&&=\frac{(2l'+1)}{8\pi}\sum_{l''} {\cal W}^{TP}_{l''} 
\left( \begin{array}{ccc}
l & l' & l'' \\
0& 0 &0 
\end{array} \right)
\left[\left( \begin{array}{ccc}
l & l' & l'' \\
-2& 2 &0 
\end{array} \right) +
\left( \begin{array}{ccc}
l & l' & l'' \\
2& -2 &0 
\end{array} \right) \right] \nonumber \\
M_{ll'}^{EE,EE}  &=& M_{ll'}^{BB,BB} \\
&&=\frac{(2l'+1)}{16\pi}\sum_{l''} {\cal W}^{PP}_{l''} 
\left[\left( \begin{array}{ccc}
l & l' & l'' \\
-2& 2 &0 
\end{array} \right) +
\left( \begin{array}{ccc}
l & l' & l'' \\
2& -2 &0 
\end{array} \right) \right]\nonumber \\
&&\times \left[\left( \begin{array}{ccc}
l & l' & l'' \\
-2& 2 &0 
\end{array} \right) +
\left( \begin{array}{ccc}
l & l' & l'' \\
2& -2 &0 
\end{array} \right) \right] \nonumber \\
M_{ll'}^{EE,BB}  &=& M_{ll'}^{BB,EE} \\
&&=\frac{(2l'+1)}{16\pi}\sum_{l''} {\cal W}^{PP}_{l''} 
\left[\left( \begin{array}{ccc}
l & l' & l'' \\
-2& 2 &0 
\end{array} \right) -
\left( \begin{array}{ccc}
l & l' & l'' \\
2& -2 &0 
\end{array} \right) \right]\nonumber \\
&&\times \left[\left( \begin{array}{ccc}
l & l' & l'' \\
-2& 2 &0 
\end{array} \right) -
\left( \begin{array}{ccc}
l & l' & l'' \\
2& -2 &0 
\end{array} \right) \right]
\end{eqnarray}
where 
\begin{equation}
{\cal W}^{ab}_l =\sum_m w^a_{lm} w^{b*}_{lm} ,
\end{equation}
with $a$ and $b$ referring to either $T$ or $P$.
All of the other coupling terms are zero. 
Note that if we use
different weighting functions for $T$, $Q$ and $U$, 
we increase the coupling between $E$ and $B$ modes.

\section{Uniform Temperature Weighting}

If we use the full sky to compute the temperature spherical harmonic terms,
then the cross-correlation term and its error matrix 
becomes particularly simple. 
For this case, 
$w_{00}^T = 1/\sqrt{4\pi}$   
and all
other coupling terms are 0.  
In this limit, the measured $c_{l'}^{TE}$ is
just a constant times the true $c_l^{TE}$
\begin{equation}
c_l^{TE} = \frac{\tilde c_{l}^{TE}} {f}
\end{equation}
where
\begin{equation}
f = \int w_E(\hat n)\frac{d\hat n}{4\pi}
\end{equation}
The covariance matrix for these terms are diagonal.
\begin{equation}
{\bf M}_{l l} = \frac{c_l^{TT} \tilde c_l^{EE} }{(2l+1) f^2}
\end{equation}

\newpage

\begin{table}
\caption{Null Tests for IQ and IU Sum and Difference Data\tablenotemark{a}
\label{sum_diff_chi_table}}
\begin{tabular}{l l c c c c c}
\tableline 
 & & & \multicolumn{2}{c}{----- Sum Map\tablenotemark{b} -----} 
     &  \multicolumn{2}{c}{----- Difference Map\tablenotemark{c} -----} \\
Correlation & Range & DOF & $\chi^2$ & Probability\tablenotemark{d}
                          & $\chi^2$ & Probability\tablenotemark{d} \\
\tableline 
IQ & $\theta < 5\arcdeg$    & 20 & 62.1  & 3$\times$10$^{-6}$     & 23.6  & 0.26 \\
IQ & $\theta \ge 5\arcdeg$  & 58 & 145.1 & 2$\times$10$^{-9}$     & 66.0  & 0.22 \\
IU & $\theta < 5\arcdeg$    & 20 & 30.9  & 0.06                   & 10.8  & 0.95 \\
IU & $\theta \ge 5\arcdeg$  & 58 & 66.1  & 0.22                   & 50.4  & 0.95 \\
\tableline
\end{tabular}

\tablenotetext{a}{$\chi^2$ comparison of the {\sl WMAP} 
correlation functions 
$C^{IQ}(\theta)$ and $C^{IU}(\theta)$ 
to a null hypothesis of CMB temperature anisotropy and instrument noise, 
but no polarization.
Temperature-polarization signals of cosmic origin
should contribute to only $C^{IQ}(\theta)$ in the sum maps.
All other tests are consistent with the null hypothesis.}

\tablenotetext{b}{Polarization sum maps 
(Q1+Q2)/2, (V1+V2)/2, (W1+W2)/2, and (W3+W4)/2
co-added with noise weights.}

\tablenotetext{c}{Polarization difference maps 
(Q1-Q2)/2, (V1-V2)/2, (W1-W2)/2, and (W3-W4)/2
co-added with noise weights.}

\tablenotetext{d}{Probability to randomly obtain $\chi^2$
larger than the measured value.}
\end{table}

\clearpage

\begin{table}
\caption{Reionization Optical Depth\tablenotemark{a}
\label{tau_table}}
\begin{tabular}{l c c c c c}
\tableline 
Data Set & Method & $\beta$ & $\tau$\tablenotemark{b} & $\chi^2$ 
         & $f(>${\sl WMAP}$)$\tablenotemark{c} \\
\tableline 

VW     &   Co-Add  &        &  $0.14^{+0.05}_{-0.03}$   &   67.0  &  0.159  \\
QVW    &   Co-Add  &        &  $0.15\pm 0.04$           &   66.2  &  0.176  \\
KaQVW  &   Co-Add  &        &  $0.14\pm 0.03$           &   97.1  &  0.001  \\
KKaQVW &   Co-Add  &        &  $0.30\pm 0.02$           &  359.8  &  0.0    \\
KKaQ   &   Co-Add  &        &  $0.29\pm 0.01$           &  476.6  &  0.0    \\
       &           &        &                           &         &         \\
QVW    &   Fit     &  -2.9  &  $0.12^{+0.19}_{-0.08}$   &  65.2   &  0.201  \\
KaQVW  &   Fit     &  -2.9  &  $0.20^{+0.14}_{-0.05}$   &  69.8   &  0.101  \\
KKaQVW &   Fit     &  -2.9  &  $0.22\pm 0.04$           &  60.9   &  0.313  \\
KKaQ   &   Fit     &  -2.9  &  $0.20\pm 0.04$           &  58.7   &  0.404  \\
       &           &        &                           &         &         \\
QVW    &   Fit     &  -3.7  &  $0.13^{+0.16}_{-0.07}$   &  66.1   &  0.180  \\
KaQVW  &   Fit     &  -3.7  &  $0.19^{+0.13}_{-0.06}$   &  68.9   &  0.117  \\
KKaQVW &   Fit     &  -3.7  &  $0.17\pm 0.03$           &  55.4   &  0.491  \\
KKaQ   &   Fit     &  -3.7  &  $0.18\pm 0.04$           &  48.0   &  0.772  \\
       &           &        &                           &         &         \\
QVW    &   Fit     &  -4.5  &  $0.13^{+0.15}_{-0.06}$   &  66.6   &  0.170  \\
KaQVW  &   Fit     &  -4.5  &  $0.15^{+0.14}_{-0.05}$   &  68.2   &  0.140  \\
KKaQVW &   Fit     &  -4.5  &  $0.16\pm 0.03$           &  57.8   &  0.419  \\
KKaQ   &   Fit     &  -4.5  &  $0.16\pm 0.04$           &  51.1   &  0.654  \\

\tableline
\end{tabular}

\tablenotetext{a}{Optical depth $\tau$ fitted from $C^{IQ}(\theta)$
for various combinations of data and foreground corrections
in a $\Lambda$CDM cosmology.
There are 57 degrees of freedom for each fit.}

\tablenotetext{b}{68\% confidence statistical uncertainties}

\tablenotetext{c}{Fraction of 1000 simulations
of reionized $\Lambda$CDM models
with $\chi^2$ larger than {\sl WMAP} value.}

\end{table}

\newpage
\clearpage

\clearpage
\begin{thebibliography}{43}
\expandafter\ifx\csname natexlab\endcsname\relax\def\natexlab#1{#1}\fi

\bibitem[{{Baccigalupi} {et~al.}(2001){Baccigalupi}, {Burigana}, {Perrotta},
  {De Zotti}, {La Porta}, {Maino}, {Maris}, \&
  {Paladini}}]{baccigalupi/etal:2001}
{Baccigalupi}, C., {Burigana}, C., {Perrotta}, F., {De Zotti}, G., {La Porta},
  L., {Maino}, D., {Maris}, M., \& {Paladini}, R. 2001, \aa, 372, 8

\bibitem[{Barnes {et~al.}(2003)}]{barnes/etal:2003}
Barnes, C. {et~al.} 2003, \apj, submitted

\bibitem[{{Becker}(2001)}]{becker/etal:2001}
{Becker}, R.~H. e.~a. 2001, \aj, 122, 2850

\bibitem[{{Bennett} {et~al.}(2003{\natexlab{a}}){Bennett}, {Bay}, {Halpern},
  {Hinshaw}, {Jackson}, {Jarosik}, {Kogut}, {Limon}, {Meyer}, {Page},
  {Spergel}, {Tucker}, {Wilkinson}, {Wollack}, \& {Wright}}]{bennett/etal:2003}
{Bennett}, C.~L., {Bay}, M., {Halpern}, M., {Hinshaw}, G., {Jackson}, C.,
  {Jarosik}, N., {Kogut}, A., {Limon}, M., {Meyer}, S.~S., {Page}, L.,
  {Spergel}, D.~N., {Tucker}, G.~S., {Wilkinson}, D.~T., {Wollack}, E., \&
  {Wright}, E.~L. 2003{\natexlab{a}}, \apj, 583, 1

\bibitem[{{Bennett} {et~al.}(2003{\natexlab{b}}){Bennett}, {Halpern},
  {Hinshaw}, {Jarosik}, {Kogut}, {Limon}, {Meyer}, {Page}, {Spergel}, {Tucker},
  {Wollack}, {Wright}, {Barnes}, {Greason}, {Hill}, {Komatsu}, {Nolta},
  {Odegard}, {Peiris}, {Verde}, \& {Weiland}}]{bennett/etal:2003b}
{Bennett}, C.~L., {Halpern}, M., {Hinshaw}, G., {Jarosik}, N., {Kogut}, A.,
  {Limon}, M., {Meyer}, S.~S., {Page}, L., {Spergel}, D.~N., {Tucker}, G.~S.,
  {Wollack}, E., {Wright}, E.~L., {Barnes}, C., {Greason}, M., {Hill}, R.,
  {Komatsu}, E., {Nolta}, M., {Odegard}, N., {Peiris}, H., {Verde}, L., \&
  {Weiland}, J. 2003{\natexlab{b}}, \apj, submitted

\bibitem[{{Bennett} {et~al.}(2003{\natexlab{c}})}]{bennett/etal:2003c}
{Bennett}, C.~L. {et~al.} 2003{\natexlab{c}}, \apj, submitted

\bibitem[{Bennett} {et~al.}(1996)]{bennett/etal:1996}
{Bennett}, C.~L.,  {Banday}, A.~J.,  {G{\' o}rski}, K.~M., {Hinshaw}, G., 
  {Jackson}, P., {Keegstra}, P.,  {Kogut}, A.,  {Smoot}, G.~F.,  
  {Wilkinson}, D.~T.,  \& {Wright}, E.~L. 1996, \apjl, 464, L1

\bibitem[{Bond \& Efstathiou(1984)}]{bond/efstathiou:1984}
Bond, J.~R. \& Efstathiou, G. 1984, \apjl, 285, L45

\bibitem[{{Brouw} \& {Spoelstra}(1976)}]{brouw/spoelstra:1976}
{Brouw}, W.~N. \& {Spoelstra}, T.~A.~T. 1976, Astron. Astrophys. Suppl. Ser.,
  26, 129

\bibitem[{{Bruscoli} {et~al.}(2002){Bruscoli}, {Tucci}, {Natale}, {Carretti},
  {Fabbri}, {Sbarra}, \& {Cortiglioni}}]{bruscoli/etal:2002}
{Bruscoli}, M., {Tucci}, M., {Natale}, V., {Carretti}, E., {Fabbri}, R.,
  {Sbarra}, C., \& {Cortiglioni}, S. 2002, New Astronomy, 7, 171

\bibitem[{{Cen}(2003)}]{cen:2003}
{Cen}, R. 2003, \apj, submitted (astro-ph/0210473)

\bibitem[{{Coulson} {et~al.}(1994){Coulson}, {Crittenden}, \&
  {Turok}}]{coulson/crittenden/turok:1994}
{Coulson}, D., {Crittenden}, R.~G., \& {Turok}, N.~G. 1994, \prl, 73, 2390

\bibitem[{{Djorgovski} {et~al.}(2001){Djorgovski}, {Castro}, {Stern}, \&
  {Mahabal}}]{djorgovski/etal:2001}
{Djorgovski}, S.~G., {Castro}, S., {Stern}, D., \& {Mahabal}, A.~A. 2001,
  \apjl, 560, L5

\bibitem[{{Fan}(2002)}]{fan/etal:2002}
{Fan}, X. e.~a. 2002, \aj, 123, 1247

\bibitem[{{Gnedin} \& {Ostriker}(1997)}]{gnedin/ostriker:1997}
{Gnedin}, N.~Y. \& {Ostriker}, J.~P. 1997, \apj, 486, 581

\bibitem[{{Haiman} {et~al.}(2000){Haiman}, {Abel}, \&
  {Rees}}]{haiman/abel/rees:2000}
{Haiman}, Z., {Abel}, T., \& {Rees}, M.~J. 2000, \apj, 534, 11

\bibitem[{{Haiman} {et~al.}(1997){Haiman}, {Rees}, \&
  {Loeb}}]{haiman/rees/loeb:1997}
{Haiman}, Z., {Rees}, M.~J., \& {Loeb}, A. 1997, \apj, 484, 985

\bibitem[{{Hinshaw} {et~al.}(2003{\natexlab{a}})}]{hinshaw/etal:2003b}
{Hinshaw}, G.~F. {et~al.} 2003{\natexlab{a}}, \apj, submitted

\bibitem[{{Hinshaw} {et~al.}(2003{\natexlab{b}})}]{hinshaw/etal:2003}
---. 2003{\natexlab{b}}, \apj, submitted

\bibitem[{{Hivon} {et~al.}(2002){Hivon}, {G{\' o}rski}, {Netterfield}, {Crill},
  {Prunet}, \& {Hansen}}]{hivon/etal:2002}
{Hivon}, E., {G{\' o}rski}, K.~M., {Netterfield}, C.~B., {Crill}, B.~P.,
  {Prunet}, S., \& {Hansen}, F. 2002, \apj, 567, 2

\bibitem[{{Hu} \& {Holder}(2003)}]{hu/holder:2003}
{Hu}, W. \& {Holder}, G.~P. 2003, \prd, submitted (astro-ph/0303400)

\bibitem[{{Hu} \& {Dodelson}(2002)}]{hu/dodelson:2002}
{Hu}, W. \& {Dodelson}, S. 2002, \araa, 40, 171

\bibitem[{{Hu} \& {Sugiyama}(1994)}]{hu/sugiyama:1994}
{Hu}, W. \& {Sugiyama}, N. 1994, \apj, 436, 456

\bibitem[{{Hu} \& {White}(1997)}]{hu/white:1997}
{Hu}, W. \& {White}, M. 1997, \prd, 56, 596

\bibitem[{{Jarosik} {et~al.}(2003)}]{jarosik/etal:2003}
{Jarosik}, N. {et~al.} 2003, \apjs, 145

\bibitem[{{Kaiser}(1983)}]{kaiser:1983}
{Kaiser}, N. 1983, \mnras, 202, 1169

\bibitem[{{Kamionkowski} {et~al.}(1997){Kamionkowski}, {Kosowsky}, \&
  {Stebbins}}]{kamionkowski/kosowsky/stebbins:1997}
{Kamionkowski}, M., {Kosowsky}, A., \& {Stebbins}, A. 1997, \prd, 55, 7368

\bibitem[{{Kaplinghat} {et~al.}(2003){Kaplinghat}, {Chu}, {Haiman}, {Holder},
  {Knox}, \& {Skordis}}]{kaplinghat/etal:2003}
{Kaplinghat}, M., {Chu}, M., {Haiman}, Z., {Holder}, G.~P.,
  {Knox}, L., \& {Skordis}, C. 2003, \apj, 583, 24

\bibitem[{{Kosowsky}(1996)}]{kosowsky:1999}
{Kosowsky}, A. 1996, Annals of Physics, 246, 49

\bibitem[{Kovac {et~al.}(2002)}]{kovac/etal:2002}
Kovac, J. {et~al.} 2002, \apj, astro-ph/0209478

\bibitem[{{Lubin} \& {Smoot}(1981)}]{lubin/smoot:1981}
{Lubin}, P.~M. \& {Smoot}, G.~F. 1981, \apj, 245, 1

\bibitem[{{Miralda-Escude}(2002)}]{miralda-escude:2003}
{Miralda-Escude}, J. 2002, \apj, submitted (astro-ph/0211071)

\bibitem[{{Oh}(2001)}]{oh:2001}
{Oh}, S.~P. 2001, \apj, 553, 499

\bibitem[{{Page} {et~al.}(2003{\natexlab{a}})}]{page/etal:2003b}
{Page}, L. {et~al.} 2003{\natexlab{a}}, \apj, submitted

\bibitem[{{Page} {et~al.}(2003{\natexlab{b}})}]{page/etal:2003}
---. 2003{\natexlab{b}}, \apj, 585, in press

\bibitem[{Peiris {et~al.}(2003)}]{peiris/etal:2003}
Peiris, H. {et~al.} 2003, \apj, submitted

\bibitem[{Rees(1968)}]{rees:1968}
Rees, M.~J. 1968, \apjl, 153, L1

\bibitem[{{Rotenberg} {et~al.}(1959){Rotenberg}, {Bivins},
{Metropolis}, \& {Wooten}}]{rotenberg/etal:1959}
{Rotenberg}, M., {Bivins}, R., {Metropolis}, N., \& {Wooten}, J. K., Jr. 
1959, The 3-j and 6-j Symbols (Cambridge: MIT)

\bibitem[{{Seljak} \& {Zaldarriaga}(1996)}]{seljak/zaldarriaga:1996}
{Seljak}, U. \& {Zaldarriaga}, M. 1996, \apj, 469, 437

\bibitem[{{Spergel} {et~al.}(2003)}]{spergel/etal:2003}
{Spergel}, D.~N. {et~al.} 2003, \apj, submitted

\bibitem[{{Tegmark}(1997)}]{tegmark/etal:1997}
{Tegmark}, M., e.~a. 1997, \apj, 474, 1

\bibitem[{{Tucci} {et~al.}(2002){Tucci}, {Carretti}, {Cecchini}, {Nicastro},
  {Fabbri}, {Gaensler}, {Dickey}, \& {McClure-Griffiths}}]{tucci/etal:2002}
{Tucci}, M., {Carretti}, E., {Cecchini}, S., {Nicastro}, L., {Fabbri}, R.,
  {Gaensler}, B.~M., {Dickey}, J.~M., \& {McClure-Griffiths}, N.~M. 2002, \apj,
  579, 607

\bibitem[{{Uyan{\i}ker} {et~al.}(1998){Uyan{\i}ker}, {Fuerst}, {Reich},
  {Reich}, \& {Wielebinski}}]{uyaniker/etal:1998}
{Uyan{\i}ker}, B., {Fuerst}, E., {Reich}, W., {Reich}, P., \& {Wielebinski}, R.
  1998, Astron. Astrophys. Suppl. Ser., 132, 401

\bibitem[{{Uyan{\i}ker} {et~al.}(1999){Uyan{\i}ker}, {Fuerst}, {Reich},
  {Reich}, \& {Wielebinski}}]{uyaniker/etal:1999}
---. 1999, Astron. Astrophys. Suppl. Ser., 138, 31

\bibitem[{{Venkatesan} {et~al.}(2003){Venkatesan}, {Tumlinson}, \&
  {Shull}}]{venkatesan/tumlinson/shull:2003}
{Venkatesan}, A., {Tumlinson}, J., \& {Shull}, J.~M. 2003, \apj, 584, 621

\bibitem[{{Venkatesan} {et~al.}(2001){Venkatesan}, {Giroux}, \&
  {Shull}}]{venkatesan/giroux/shull:2001}
{Venkatesan}, A., {Giroux}, M.~L., \& {Shull}, J.~M. 2001, \apj, 563, 1

\bibitem[{{Wyithe} \& {Loeb}(2003)}]{wyithe/loeb:2003}
{Wyithe}, J.~S.~B. \& {Loeb}, A. 2003, \apj, 586, 693

\bibitem[{{Zaldarriaga}(1997)}]{zaldarriaga:1997}
{Zaldarriaga}, M. 1997, \prd, 55, 1822

\bibitem[{{Zaldarriaga} \& {Seljak}(1997)}]{zaldarriaga/seljak:1997}
{Zaldarriaga}, M. \& {Seljak}, U. 1997, \prd, 55, 1830

\end{thebibliography}
\end{document}